\documentclass{aa}  

\usepackage{graphicx}
\usepackage{subcaption}
\usepackage{txfonts}
\usepackage{hyperref}
\usepackage{xcolor}
\usepackage{tikz,xcolor}
\usepackage{csquotes}
\usepackage{siunitx}

\usepackage{longtable}
\usepackage[para]{threeparttable}
\usepackage[para]{threeparttablex}

\usepackage{scalerel}
\usepackage{tikz}
\usetikzlibrary{svg.path}

\definecolor{orcidlogocol}{HTML}{A6CE39}
\tikzset{
  orcidlogo/.pic={
    \fill[orcidlogocol] svg{M256,128c0,70.7-57.3,128-128,128C57.3,256,0,198.7,0,128C0,57.3,57.3,0,128,0C198.7,0,256,57.3,256,128z};
    \fill[white] svg{M86.3,186.2H70.9V79.1h15.4v48.4V186.2z}
                 svg{M108.9,79.1h41.6c39.6,0,57,28.3,57,53.6c0,27.5-21.5,53.6-56.8,53.6h-41.8V79.1z M124.3,172.4h24.5c34.9,0,42.9-26.5,42.9-39.7c0-21.5-13.7-39.7-43.7-39.7h-23.7V172.4z}electronically
                 svg{M88.7,56.8c0,5.5-4.5,10.1-10.1,10.1c-5.6,0-10.1-4.6-10.1-10.1c0-5.6,4.5-10.1,10.1-10.1C84.2,46.7,88.7,51.3,88.7,56.8z};
  }
}

\newcommand\orcidicon[1]{\href{https://orcid.org/#1}{\mbox{\scalerel*{
\begin{tikzpicture}[yscale=-1,transform shape]
\pic{orcidlogo};
\end{tikzpicture}
}{|}}}}

\newcommand{\review}[1]{#1} 
\newcommand{\reviewtwo}[1]{#1}
\begin{document}

   \title{TELAMON: Effelsberg monitoring of AGN jets with very-high-energy astroparticle emission}

   \subtitle{I. Program description and sample characterization}
   \author{F.\,Eppel\inst{1,2}
          \and
          M.\,Kadler\inst{1}
          \and
          J.\,Heßdörfer\inst{1,2}
          \and
          P.\,Benke\inst{2,1}
          \and
          L.\,Debbrecht\inst{2}
          \and
          J.\,Eich\inst{1}
          \and
          A.\,Gokus\inst{3,4,1}
          \and
          S.\,Hämmerich\inst{4}
          \and
          D.\,Kirchner\inst{1}
          \and
          G.\,F.\,Paraschos\inst{2}
          \and
          F.\,Rösch\inst{1,2}
          \and
          W.\,Schulga\inst{1}
          \and
          J.\,Sinapius\inst{5}
          \and
          P.\,Weber\inst{1}
          \and
          U.\,Bach\inst{2}
          \and
          D.\,Dorner\inst{6,1}
          \and
          P.\,G.\,Edwards\inst{7}
          \and
          \review{M.\,Giroletti\inst{8}}
          \and
          A.\,Kraus\inst{2}
          \and
          O.\,Hervet\inst{9}
          \and
          S.\,Koyama\inst{10,11}
          \and 
          T.\,P.\,Krichbaum\inst{2}
          \and
          K.\,Mannheim\inst{1}
          \and
          E.\,Ros\inst{2}
          \and
          M.\,Zacharias\inst{12,13}
          \and
          J.\,A.\,Zensus\inst{2}
          }

   \institute{Julius-Maximilians-Universität Würzburg, Institut für Theoretische Physik und Astrophysik, Lehrstuhl für Astronomie, Emil-Fischer-Straße 31, D-97074 Würzburg, Germany\\
              \email{florian.eppel@uni-wuerzburg.de}
         \and
         Max-Planck-Institut für Radioastronomie, Auf dem Hügel 69, D-53121 Bonn, Germany
        \and
        Department of Physics \& McDonnell Center for the Space Sciences, Washington University in St. Louis, One Brookings Drive, St. Louis, MO-63130, USA
        \and
        Remeis Observatory and Erlangen Centre for Astroparticle Physics, Universität Erlangen-Nürnberg, Sternwartstr. 7, \reviewtwo{D-}96049 Bamberg, Germany
        \and
        Deutsches Elektronen-Synchrotron DESY, Platanenallee 6, \reviewtwo{D-}15738 Zeuthen, Germany
        \and
        ETH Zürich, CH-8093 Zürich, Switzerland
        \and
         CSIRO Space \& Astronomy, ATNF, PO Box 76. Epping NSW 1710. Australia
         \and
         \review{INAF Istituto di Radioastronomia, via Gobetti 101, \reviewtwo{I-}40129 Bologna, Italy}
         \and
         Santa Cruz Institute for Particle Physics and Department of Physics, University of California, Santa Cruz, CA 95064, USA
         \and
         Graduate School of Science and Technology, Niigata University, 8050 Ikarashi 2-no-cho, Nishi-ku, Niigata 950-2181, Japan
         \and
         Institute of Astronomy and Astrophysics, Academia Sinica, 11F of Astronomy-Mathematics Building, AS/NTU No. 1, Sec. 4, Roosevelt Rd, Taipei 10617, Taiwan, R.O.C.
         \and
         Landessternwarte, Universität Heidelberg, Königsstuhl, D-69117 Heidelberg, Germany
         \and
         Center for Space Research, North-West University, Potchefstroom 2520, South Africa
        }

   \date{Received October 13, 2023; accepted January 11, 2024}

 
  \abstract
   {}
   {We introduce the TELAMON program which is using the Effelsberg 100-m telescope to monitor the radio spectra of active galactic nuclei (AGN) under scrutiny in astroparticle physics, specifically TeV blazars and candidate neutrino-associated AGN. Here, we present and characterize our main sample of TeV-detected blazars.}
   {We analyze the data sample from the first $\sim$2.5 years of observations between August 2020 and February 2023 in the range from 14\,GHz to 45\,GHz. During this pilot phase, we have observed all 59 TeV-detected blazars in the Northern Hemisphere (i.e., Dec. $>0^\circ$) known at the time of observation. We discuss the basic data reduction and calibration procedures used for all TELAMON data and introduce a sub-band averaging method used to calculate average light curves for the sources in our sample.}
   {\reviewtwo{T}he TeV-selected sources in our sample exhibit a median flux density of 0.12\,Jy at 20\,mm, 0.20\,Jy at 14\,mm and 0.60\,Jy at 7\,mm. The spectrum for most of the sources is consistent with a flat radio spectrum and we find a median spectral index ($S(\nu)\propto\nu^\alpha$) of $\alpha=-0.11$. Our results on flux density and spectral index are consistent with previous studies of TeV-selected blazars. Compared to the GeV-selected F-GAMMA sample, TELAMON sources are significantly fainter in the radio band. This is consistent \review{with the double-humped spectrum of blazars being shifted towards higher frequencies for TeV-emitters (in particular for \reviewtwo{high-synchrotron peaked BL Lac type objects}), which results in a lower radio flux density.} 
   \reviewtwo{The spectral index distribution of our} TeV-selected blazar sample \review{is not significantly different from} the GeV-selected F-GAMMA sample. 
   Moreover, we present a strategy to track the light curve evolution of sources in our sample for future variability and correlation analysis.}
   {}


\authorrunning{F.\,Eppel et al.}

\maketitle
%
\section{Introduction}
Blazars are radio-loud active galactic nuclei (AGN) hosting relativistic jets pointed close to our line of sight. Their emission is highly beamed and Doppler-boosted, which makes them variable broadband emitters from radio to $\gamma$-ray energies. 
Their spectral energy distribution (SED) typically shows a two-peaked (double-humped) spectrum. The first peak corresponds to synchrotron emission while the second peak is often attributed to inverse Compton scattering. High-synchrotron peaked BL Lac type objects (HBLs) are defined as sources whose primary (synchrotron) emission hump peaks above $10^{15}$\,Hz in $\nu F_\nu$ scale \citep{Padovani1995}. In the most extreme cases, they peak at even higher frequencies by up to two orders of magnitude for the so called extreme HBLs \citep[EHBLs,][]{Ghisellini1999,Biteau2020}. The second, high-energy peak of blazar SEDs can stretch into the very-high-energy (VHE) regime at TeV $\gamma$-rays. Imaging Air Cherenkov Telescopes have been able to detect VHE emission from $\sim$80 blazars, with the majority of them being HBLs\footnote{\label{tevcatnote}\url{http://tevcat2.uchicago.edu}}. Most of these sources are faint radio sources, which makes TeV-emitting blazars difficult to study in the radio band.
Blazars are of utmost interest for astroparticle physics as potential sources of ultrahigh-energy cosmic rays and neutrinos \citep[e.g.,][]{Hillas1984,Mannheim1995}. In particular, HBLs and EHBLs have been considered in some recent theoretical works as relevant neutrino sources \citep{Tavecchio2014,Padovani2015,Giommi2020}. We have established the TeV Effelsberg Long-term AGN Monitoring (TELAMON) program in August 2020, a radio monitoring program which uses the Effelsberg 100-m telescope to investigate the radio properties of TeV-emitting blazars. We perform radio observations of a sample of TeV-selected blazars at high frequencies up to 44\,GHz to trace dynamical process in these objects. Our program is designed to monitor the radio spectra of TeV-blazars and candidate neutrino AGN.
A first study of TeV-selected HBLs and EHBLs has been presented by \cite{Lindfors} at 15 GHz. We extend their observations with radio coverage at multiple frequencies, and, for the first time, a spectral characterization of these objects. In Section\,\ref{sec:obs}, we explain in detail the sample selection and observing setup used with the Effelsberg 100-m telescope. On top of that, we present the TELAMON analysis pipeline including cross-scan setup and calibration procedures. In Section\,\ref{sec:results}, we present first results for sources in our sample and present average flux densities and spectral indices for all observed sources. These results are discussed and compared with previous studies, e.g., the \cite{Lindfors} study, in Section\,\ref{sec:discussion}. Moreover, we compare source properties of the TeV-selected sources in our sample with the GeV-selected F-GAMMA \citep{Fuhrmann,Angelakis2019} sample and discuss similarities and differences. In Section \ref{sec:conclusions}, we give an overview about planned future publications and the overall relevance of the TELAMON project within current developments in astroparticle physics.

\section{Observations \& analysis}
\label{sec:obs}
\subsection{Program description \& sample}

Our sample consists of all 59 TeV-detected blazars\footref{tevcatnote} in the Northern Hemisphere (i.e., Dec. $>0^\circ$). In addition, we include 5 TeV-detected blazars from the Southern Hemisphere and a TeV-detected radio galaxy (3C\,264) as sources of special interest. This TeV-selected sample, consisting of 65 sources, is presented in Table~\ref{tab:targetlist}. It includes \review{16 EHBLs (i.e., $\nu_\mathrm{peak}>\,10^{17.38}\,\mathrm{Hz}\, \widehat{=}\, 1$\,keV), 32 HBLs (i.e., $10^{15}\,\mathrm{Hz}<\nu_\mathrm{peak}\leq10^{17.38}$\,Hz), 4 intermediate-peaked BL Lac type objects (IBL, i.e., $10^{14}\,\mathrm{Hz}<\nu_\mathrm{peak}\leq10^{15}$\,Hz), 4 low-peaked BL Lac type objects (LBL, i.e., $\nu_\mathrm{peak}\leq10^{14}$\,GHz)}, 8 flat-spectrum radio quasars (FSRQ) and 1 Radio Galaxy (RG). For J0316+4119 (IC\,310), a clear \review{classification is complicated}  since the source shows some blazar-like properties \review{\citep{Kadler2012}} but its jet might be misaligned by 10$-$20$^\circ$ from the line of sight \review{\citep{angleIC310}}.
We include the source in our HBL count, according to its peak frequency found in the literature. Note that \review{our classification for BL Lac type objects solely depends on the synchrotron peak frequency, as per the references listed in Table~\ref{tab:targetlist}. We adapt the extreme synchrotron limit (i.e., $\nu_\mathrm{peak}>1$\,keV) from \cite{Biteau2020}, but do not account for the other types of extreme behaviours introduced in their work. Therefore, in our analysis, some of the  blazars labeled as `extreme' by \cite{Biteau2020} fall into the HBL category in our work. On the other hand, not all EHBL sources from our analysis are covered in \cite{Biteau2020}.}

\begin{table*}
\caption{\small The TELAMON sample of TeV-emitting AGN. The classification into EHBL, HBL, IBL and LBL is performed according to the synchrotron peak frequency found in the literature.}
\centering
\footnotesize
\resizebox{17.5cm}{!}{
\begin{tabular}{@{}c@{\,}c@{\,}c@{\,}c@{~~}c@{~~~}c@{\,}c@{}}
\hline\hline
{ID}  & {Alternative} & {Class} & {Synchr. Peak}$^\textrm{a}$ & {Redshift$^\textrm{b}$} & {Synchr. Peak}  & {Redshift} \\
  {(J2000)}   &  {Name}  & {} & {log$_{10}\left(\frac{\nu_\mathrm{peak}}{\mathrm{Hz}}\right)$}  & {} & {Reference} & {Reference} \\
\hline
J0035+5950 & 1ES\,0033+595 & EHBL &  18.\reviewtwo{4} & 0.\review{467} & \cite{3HSP}  & \review{\cite{redshift0035}} \\
J0112+2244 & S2\,0109+22 & IBL &  \reviewtwo{14.41} & \review{0.265} & \reviewtwo{\cite{Zhou2021}}  & \review{\cite{redshift0109}} \\
J0136+3906 & RGB\,J0136+391 & HBL & \reviewtwo{16.27} & - & \reviewtwo{\cite{Fan2016}} & - \\
J0152+0146* & RGB\,J0152+017 & \review{HBL} & \reviewtwo{16.25} & 0.080 & \reviewtwo{\cite{Zhou2021}} & \review{\cite{redshift0152}} \\
J0214+5144* & TXS\,0210+515 & \review{HBL} & 17.3 & 0.049 & \cite{3HSP}  & \review{\cite{redshift0214}} \\
J0221+3556 & S3\,0218+35 & FSRQ &  \reviewtwo{13.98} & \review{0.954} & \reviewtwo{\cite{Zhou2021}}  & \review{\cite{redshift0221}} \\
J0222+4302 & 3C\,66A & HBL &  15.\reviewtwo{76} & 0.34\review{0} & \cite{PeakFreq0521} & \review{\cite{redshift0222}} \\
J0232+2017* & 1ES\,0229+200 & EHBL &  18.\reviewtwo{6} & 0.140 & \cite{3HSP}   & \review{\cite{redshift0232}} \\
J0303$-$2407 & PKS 0301$-$243 & HBL &  15.\reviewtwo{8} & \review{0.263} & \cite{3HSP}  & \review{\cite{redshift0303}} \\
J0316+4119 & IC\,310 & \review{RG/HBL} & 17 & \review{0.0190} & \cite{PeakFreqIC310} & \review{\cite{redshift0316}} \\
J0319+1845 & RBS 0413 & \review{HBL} & \reviewtwo{16.37} & 0.190 & \reviewtwo{\cite{Zhou2021}}  & \review{\cite{redshift0319}} \\
J0416+0105* & 1ES\,0414+009 & \review{HBL} & 16.\reviewtwo{6} & 0.287 & \cite{3HSP}  & \review{\cite{redshift0416}} \\
J0507+6737 & 1ES\,0502+675 & EHBL & \reviewtwo{18.0}  & 0.34\review{0} & \cite{3HSP}  & \review{\cite{redshift0507}} \\
J0509+0541 & TXS\,0506+056 & HBL &  15.\reviewtwo{47} & 0.3365 & \cite{PeakFreq0521}  & \review{\cite{redshift0509}} \\
J0521+2112 & RGB\,J0521+212 & \review{HBL} &  \reviewtwo{$>$15.2} & \review{$>$0.18}  & \cite{3HSP}  & \review{\cite{redshift0035}}\\
J0648+1516 & RX J0648.7+1516 & HBL & \reviewtwo{16.71} & 0.179 & \reviewtwo{\cite{Zhou2021}}  & \review{\cite{Aliu2011}} \\
J0650+2502  & 1ES\,0647+250 & HBL &  16.\reviewtwo{8} & \review{0.41}  & \cite{3HSP} & \review{\cite{redshift0650}} \\
J0710+5909* & RGB\,J0710+591 & EHBL & 18.\reviewtwo{2} & 0.12\review{5} & \cite{3HSP}  & \review{\cite{redshift0710}} \\
J0721+7120 & S5\,0716+714 & IBL & \reviewtwo{14.95} & \reviewtwo{0.2304}$^\ddagger$ & \reviewtwo{\cite{Zhou2021}} & \reviewtwo{\cite{Pichel2023}}  \\
J0733+5153* & PGC 2402248 & EHBL & 17.9 & 0.0\review{650} & \cite{3HSP}  & \review{\cite{redshift0733}} \\
J0739+0136 & [HB89] 0736+017 & FSRQ & \reviewtwo{13.97} & 0.18941 & \reviewtwo{\cite{Zhou2021}} & \cite{Grasha2019}  \\
J0809+5219 & 1ES 0806+524 & HBL & \reviewtwo{15.95}  & 0.138 & \reviewtwo{\cite{Zhou2021}} & \review{\cite{redshift0809}} \\
J0812+0237 & 1RXS J081201.8+023735 & EHBL & \reviewtwo{$>$}17.5 & 0.17\review{3} & \cite{0812paper}  & \review{\cite{redshift0812}} \\
J0847+1133* & RBS 0723 & EHBL & 17.\reviewtwo{9} & 0.198  & \cite{3HSP} & \review{\cite{redshift0847}} \\ 
J0854+2006$\dagger$ & OJ287 & LBL & \reviewtwo{13.29}  & 0.306 & \reviewtwo{\cite{Zhou2021}}  & \review{\cite{redshift0854}} \\
J0913$-$2103 & MRC\,0910$-$208 & \review{HBL} & 17.\reviewtwo{2}  & 0.198 & \cite{3HSP}  & \review{\cite{redshift0913}} \\
J0958+6533 & S4 0954+658 & FSRQ & \reviewtwo{12.98} & 0.3694 & \reviewtwo{\cite{Zhou2021}} & \cite{Gonzalez20221} \\
J1015+4926 & 1ES\,1011+496 & HBL &  16.\reviewtwo{5} & 0.2\review{12} & \cite{3HSP}  & \review{\cite{redshift1015}} \\
J1058+2817 & GB6\,J1058+2817 & EHBL &  18.\reviewtwo{4}7 & 0.254 & \cite{PeakFreq0521}  & \review{\cite{redshift1058}} \\
J1104+3812* & Mrk\,421 & \review{HBL} &  16.3 &  0.03\review{08} & \cite{3HSP}  & \review{\cite{redshift1104}} \\
J1136+7009 & Mrk\,180 & HBL &  16.8 & 0.045\review{8} & \cite{3HSP}  & \review{\cite{redshift1136+7009}}  \\
J1136+6737* & RX J1136.5+6737 & EHBL & 18.\reviewtwo{2} & 0.134 & \cite{3HSP} & \review{\cite{redshift1136+6737}} \\
J1145+1936 & 3C\,264 & RG & \reviewtwo{$\sim17$} & 0.0216 & \reviewtwo{\cite{SED1145}}& \cite{SDSS} \\
J1159+2914 & 4C+29.45 & FSRQ & \reviewtwo{13.05} & 0.724745 & \reviewtwo{\cite{Zhou2021}} & \cite{Albareti2017} \\
J1217+3007 & ON\,325 & HBL &  15.\reviewtwo{63} & 0.1\review{29} & \cite{PeakFreq0521}  & \review{\cite{redshift0035}} \\
J1221+3010* & 1ES\,1218+304 & \review{HBL} & 16.\reviewtwo{9} & 0.18\review{4} & \cite{3HSP}  & \review{\cite{redshift0809}} \\
J1221+2813 & W\,Comae & IBL &  14.8\reviewtwo{8} & 0.102 & \cite{PeakFreq0521}  & \review{\cite{redshift1221+2813}} \\
J1224+2436 & MS 1221.8+2452 & HBL & 15.68 & 0.219 & \cite{Zhou2021}  & \review{\cite{redshift1224+2436}} \\
J1224+2122 & PKS\,1222+21 & FSRQ & \reviewtwo{13.87} & 0.433826 & \reviewtwo{\cite{Zhou2021}} & \cite{Albareti2017} \\
J1230+2518 & ON\,246 & IBL &  \reviewtwo{14.75} & \review{0.555}  & \reviewtwo{\cite{Zhou2021}}  & \review{\cite{redshift1230}} \\
J1415+1320 & PKS\,1413+135 & LBL & \reviewtwo{12.96} & \review{0.334} & \reviewtwo{\cite{PeakFreq0521}}  & \review{\cite{redshift1415}} \\
J1422+3223 & OQ\,334 & FSRQ &  \reviewtwo{$<15$} & 0.68\review{2} & \reviewtwo{\cite{SED1422}} & \review{\cite{redshift1422}} \\
J1427+2348 & OQ\,240 & HBL &  15.\reviewtwo{9} & 0.6\review{05} & \cite{PeakFreq0521}  & \review{\cite{redshift0035}} \\
J1428+4240* & 1ES\,1426+428 & EHBL &  18.\reviewtwo{2} & 0.129  & \cite{3HSP}  & \review{\cite{redshift1428}} \\
J1443+1200* & 1ES\,1440+122 & EHBL & 17.\reviewtwo{8} & 0.16\review{3}  & \cite{3HSP} & \review{\cite{redshift0232}}  \\
J1443+2501 & PKS\,1441+25 & FSRQ &  \reviewtwo{13.39} & 0.939  & \reviewtwo{\cite{Zhou2021}} & \cite{RedshiftPKS1441} \\
J1518$-$2731 & TXS\,1515$-$273 & HBL &  15.\reviewtwo{4} & 0.1\review{28} &  \cite{3HSP}  & \review{\cite{Gonzalez20221}} \\
J1555+1111 & PG\,1553+113 & HBL &  15.6 & \review{0.028} & \cite{3HSP}  & \review{\cite{redshift1555}} \\
J1653+3945* & Mrk\,501 & EHBL &  17.9 & 0.03\review{35} & \cite{3HSP}  & \review{\cite{redshift1422}} \\
J1725+1152 & H 1722+119 & HBL & 16.46 & 0.\review{028} & \cite{Zhou2021}  & \review{\cite{redshift1555}} \\
J1728+5013* & I\,Zw\,187 & \review{HBL} &  17.0 & 0.055\review{4} & \cite{3HSP}  & \review{\cite{redshift1728}} \\
J1743+1935* & 1ES\,1741+196 & EHBL &  17.8 & 0.08\review{4} & \cite{3HSP}  & \review{\cite{redshift1743}} \\
J1751+0938 & PKS\,1749+096 & LBL & 13.\reviewtwo{23} & 0.32 & \reviewtwo{\cite{Zhou2021}} & \review{\cite{redshift1751}} \\
J1813+3144 & B2\,1811+31 & FSRQ & 15.0  & 0.117 & \cite{3HSP}  & \review{\cite{redshift0710}} \\
J1943+2118 & HESS J1943+213 & EHBL & 18.\reviewtwo{2} & 0.2 & \cite{3HSP}  & \review{\cite{redshift1943}}\\
J1958$-$3011 & 1RXS\,J195815.6$-$301119 & \review{HBL} &  17.0 & 0.119  & \cite{3HSP} & \review{\cite{redshift1958}} \\
J1959+6508* & 1ES\,1959+650 & \review{HBL} &  16.9 & 0.047\review{0}  & \cite{3HSP} & \review{\cite{redshift0232}} \\
J2001+4352 & MAGIC\,J2001+435 & HBL & $\sim$16 & 0.18  & \cite{MAGIC2001} & \review{\cite{MAGIC2001}} \\
J2039+5219* & 1ES\,2037+521 & HBL & 16.2 & 0.053 & \cite{Zhou2021}  & \review{\cite{redshift2039}} \\
J2056+4940 & RGB\,J2056+496 & EHBL & 17.6 & \review{-} & \cite{3HSP}  & \review{-} \\
J2158$-$3013 & PKS\,2155$-$304 & HBL &  15.4 & 0.11\review{6}  & \cite{3HSP} & \review{\cite{redshift2158}} \\
J2202+4216 & BL Lac & LBL & \reviewtwo{12.95} & 0.06\review{86} & \reviewtwo{\cite{Zhou2021}} & \review{\cite{redshift2202}} \\
J2243+2021 & RGB J2243+203 & \review{HBL} &  15.\reviewtwo{3} & \review{0.53} & \cite{3HSP}  & \review{\cite{redshift2243}} \\
J2250+3825 & B3\,2247+381 & HBL & 16.33 & 0.119 & \cite{Zhou2021}  & \review{\cite{redshift0152}} \\
J2347+5142* & 1ES\,2344+514 & EHBL &  17.7 & 0.044 & \cite{3HSP}  & \review{\cite{redshift2347}}  \\
\hline
\multicolumn{7}{l}{\footnotesize \reviewtwo{$^\textrm{a}$ Synchrotron peak frequency in the source rest frame, i.e., $\nu_{\textrm{peak,rest}}=\nu_{\textrm{peak,obs}}(1+z)$. Already corrected values from the literature were adjusted}} \\
\multicolumn{7}{l}{\footnotesize 
\reviewtwo{to the redshift values in this work.}
$^\textrm{b}$For some sources no \review{conclusive redshift value is available in the literature. See also \cite{Foschini2022}}} \\
\multicolumn{7}{l}{\footnotesize \review{for a more detailed collection of redshifts for most sources including information about their reliability. *Sources overlapping with }} \\
\multicolumn{7}{l}{\cite{Biteau2020}. $\dagger$ Observations are coordinated with the MOMO program \citep{Komossa}. \reviewtwo{$^\ddagger$ The adopted
redshift is a statistical}} \\
\multicolumn{7}{l}{\reviewtwo{estimate, which may deviate from the $z=0.31$, previously suggested based on marginal detection of the host galaxy by \cite{Nilsson2008}.}} 
\end{tabular}
}
\label{tab:targetlist}
\end{table*}

\subsection{Observations}
\label{sec:observingprocess}
For our observations, we are using the Effelsberg 100-m telescope operated by the Max-Planck-Institute for Radio Astronomy in Bonn, Germany. 
The observations presented here are conducted with the 20\,mm, 14\,mm and 7\,mm receivers mounted in the secondary focus, in continuum observing mode with the backend \enquote{dual-spec-OPTOCBE}. This results in four sub-bands (centered at 19.25\,GHz, 21.15\,GHz, 22.85\,GHz and 24.75\,GHz) for the 14\,mm receiver, and four sub-bands (centered at 36.25\,GHz, 38.75\,GHz, 41.25\,GHz, 43.75\,GHz) for the 7\,mm receiver, which we are using since August 2020. Since spring 2021, the 20\,mm receiver has been added yielding two additional frequency bands (centered at 14.25\,GHz and 16.75\,GHz). Note that we also started using the \enquote{SpecPol} receiver backend to record polarization information at the same time \citep[cf.][]{TELAMONpolarization}. The polarization analysis and first results will be assessed in a separate publication. All receivers are equipped with two horns. The first horn is pointed directly at the target and passes the signal directly to the receiver. The second horn is pointed at the atmosphere off-source and is used to subtract weather effects from the first horn. We apply this weather subtraction to all of our observations, because at the used frequencies Earth's atmosphere (especially on cloudy days) emits thermal radiation that needs to be accounted for.
In order to measure the flux density of a source, \enquote{cross-scans} are performed on the targets, consisting of typically 8 sub-scans (4 in azimuth and 4 in elevation) at 20\,mm, 14\,mm and 7\,mm (36.25\,GHz, 38.75\,GHz). For the higher 7\,mm frequencies (41.25\,GHz, 43.75\,GHz), 16 sub-scans (8 in azimuth and 8 in elevation) are used. During a cross-scan the telescope slews over the point-like source region in azimuth- and elevation-direction multiple times while measuring the antenna temperature of the receiver. About every four hours, a calibrator source is observed in order to focus the telescope and to extract calibration factors (cf. Sect.~\ref{sec:calibration}).

For this work, we consider all observations from within the first 2.5\,years of the program, i.e., from August 2020 to February 2023. This comprises data taken in 95 observing sessions (epochs) with a total observing time of 1160.3\,hours. Note that part of the observing time was also used for observations of neutrino-candidate sources \review{\citep[cf.][]{TELAMON}}, which will be discussed in a separate publication. During this time, we have observed 65 distinct TeV-sources in total, 53 at 20\,mm, 61 at 14\,mm and 45 at 7\,mm. \review{The number of observations per source and per receiver varies within the sample since we have tested different observing strategies depending on the source flux density. 
Our optimized monitoring strategy 
has been
implemented since mid-2022 (cf. Sec.\,\ref{sec:discussion_detectrates}).}
In order to select the statistically most complete sample from our observed targets, we will from here on only address the 59 Northern (i.e., Dec. $>0^\circ$) TeV-blazars. Results for the 5 Southern sources and the radio galaxy 3C\,264 (J1145+1936) are presented in Appendix \ref{app:extra_sources}. Plots of the latest available light curves and spectra are publicly available on the dedicated TELAMON website\footnote{\url{https://telamon.astro.uni-wuerzburg.de/}}.

\subsection{Data acquisition \& reduction}
The general data analysis for pointed flux density measurements with the Effelsberg 100-m telescope has been described in detail by \cite{Angelakis2019} in context of the F-GAMMA program. We use a very similar data reduction procedure and will therefore restrict this section to the changes and improvements compared to the analysis of \cite{Angelakis2019}. The major improvements presented here are a semi-automated flagging algorithm, a new calibration procedure, and an in-depth measurement uncertainty discussion. In principle, this section explains how the raw data output of every scan, namely antenna temperatures, is converted into astrophysical units of jansky. The antenna temperatures are calculated by using a noise diode switching system \citep[cf.][]{MuellerEffelsberg}. Left-handed circular polarization (LCP) and right-handed circular polarization (RCP) are averaged.

\subsubsection{Sub-scan fitting}
\label{sec:subscan_fitting}

As mentioned in Sect.~\ref{sec:observingprocess}, every scan of a source at a specific frequency consists of multiple sub-scans. These sub-scans are used to calculate one flux density value for each frequency per scan. For every scan, all azimuth sub-scans are used to generate a single average azimuth scan and all elevation sub-scans are used to generate a single average elevation scan. Since all of the sample sources are assumed to be point-like and the primary beam/point spread function (PSF) of the telescope is well described by a 2D Gaussian, a Gaussian curve is fitted to both average scans. This fitting process is carried out with the \texttt{toolbox} software \citep{Kraus2003}. At this point, it is necessary to perform a data quality check to see if all sub-scans can be considered clean scans. This is important, since some sub-scans might be corrupted by radio frequency interference (RFI), telescope errors, or atmospheric effects, and therefore might lead to errors in the derived flux density values.

In order to filter out corrupted sub-scans, a flagging system has been developed, which uses multiple criteria to detect corrupted scans. The detailed flagging criteria are listed in Appendix \ref{app:flagging_criteria}. If an averaged scan is flagged according to these criteria, it is not trivial to judge whether the source was below the detection limit (not visible in all sub-scans) or if there were simply a few corrupted sub-scans without checking every single sub-scan for every source. As done by \cite{Angelakis2019} a manual analysis of individual sub-scans is possible but very tedious, time-consuming, and also not clearly reproducible since it depends on the analyzer's individual eyes. Therefore, we have developed a semi-automated analysis tool which takes care of detecting and sorting-out corrupted sub-scans. The general principle of the algorithm is presented in Appendix\,\ref{app:flagging_criteria}. The automated analysis quality is superior to the manual analysis since every sub-scan deletion can be tried out, instead of simply judging by the human eye what sub-scan deletion might improve the overall fit quality. On top of that, due to the appointed flagging criteria, it is completely reproducible. As a final check, the data are again inspected manually to sort out any outliers and left-over corrupted scans, which the algorithm was not able to detect.

\subsubsection{Data corrections}
\label{sec:pointcorr}
All averaged scans that passed our semi-automated flagging algorithm undergo further data reduction processes. These data reduction are composed of the pointing offset correction, atmospheric opacity correction, and elevation-dependent gain correction presented by \cite{Angelakis2019} in their Sect.\,4. The only difference is that we determine the zenith opacity $\tau$ using a water-vapor radiometer located at the focus cabin of the telescope by measuring the strength of the 22\,GHz water-vapor line \citep{wvrEff}. On top of that, we use updated gain curve parameters $A_\mathrm{i}$ as presented in Table\,\ref{tab:gaincurve}.

\begin{table}[]
    \centering
    \caption{Gain curve parameters used for the different receivers of the Effelsberg 100-m telescope \review{\citep[\reviewtwo{second order polynom,} cf.~Eq.\,(5) of][]{Angelakis2019}}.}
    \begin{tabular}{ccccc}
    \hline\hline
        Receiver & $A_0$ & $A_1 \times 10^{3}$ & $A_2 \times 10^5$ \\
        \hline
        20\,mm & 0.971 & 1.833 & $-$2.867 \\
        14\,mm & 0.962 & $-$1.950 & $-$2.493 \\
        7\,mm (36$-$39\,GHz) & 0.835 & 8.312 & $-$10.46 \\
        7\,mm (41$-$44\,GHz) & 0.785 & 11.26 & $-$14.75 \\
        \hline
    \end{tabular}
    \label{tab:gaincurve}
\end{table}

\subsection{Calibration}
\label{sec:calibration}
\subsubsection{Calibrator sources}
In this section, it is discussed how the calibration factor for the analysis is determined. \cite{Baars1977} have introduced a set of secondary calibrators that can be used on a day-to-day basis for calibration purposes. Their flux density is very constant over long periods of time (several decades) and they are also very compact sources. These secondary calibrators have been further investigated and monitored by \cite{Ott1994} and most lately by \cite{Pearly2013,Pearly2017}. According to the latest publication \citep{Pearly2017}, the best suited calibrators in the TELAMON frequency range (14\,GHz$-$44\,GHz) are 3C\,286 and 3C\,295. They are therefore used as calibrator sources in this work. \cite{Pearly2017} provide parametrized spectra for these sources that are used to calculate the frequency dependent calibrator flux densities for our analysis. Since 3C\,295 is quite faint ($\lesssim 1\,$Jy) at higher frequencies ($\nu\gtrsim35$\,GHz) and therefore not always detectable (especially during bad weather sessions), the sources NGC\,7027 and W3(OH) are also included as calibrator sources. NGC\,7027 is a planetary nebula and has been proposed as a secondary calibrator by \cite{Baars1977}. Even though its flux density is gradually fading with time, this source can be used as a calibrator since its behavior has been well characterized by \cite{Zijlstra2008}. The spectral model provided in this paper is used to calculate time-dependent calibration flux densities for NGC\,7027. W3(OH) is a star forming region that exhibits a strong water maser in the 14\,mm band \citep{W3OH}. Therefore, it is excluded for 14\,mm calibration, but since the source is very bright at 7\,mm ($\approx 3$\,Jy) it is useful as a substitute for 3C\,295 in this band. For W3(OH) we use our own calibration model which was created using Effelsberg archival data and assuming a free-free emission model. Due to their brightness, the two main calibrators observed with highest priority are 3C\,286 and NGC\,7027. 3C\,295 is used as a backup calibrator at 20\,mm and 14\,mm, and W3(OH) as a backup calibrator at 20\,mm and 7\,mm. For one epoch in 2021, we used 3C\,138 as a calibrator for all bands, using the \cite{Pearly2017} model.

\subsubsection{Calibration process}

\begin{figure}
    \centering
    \includegraphics[width=\columnwidth]{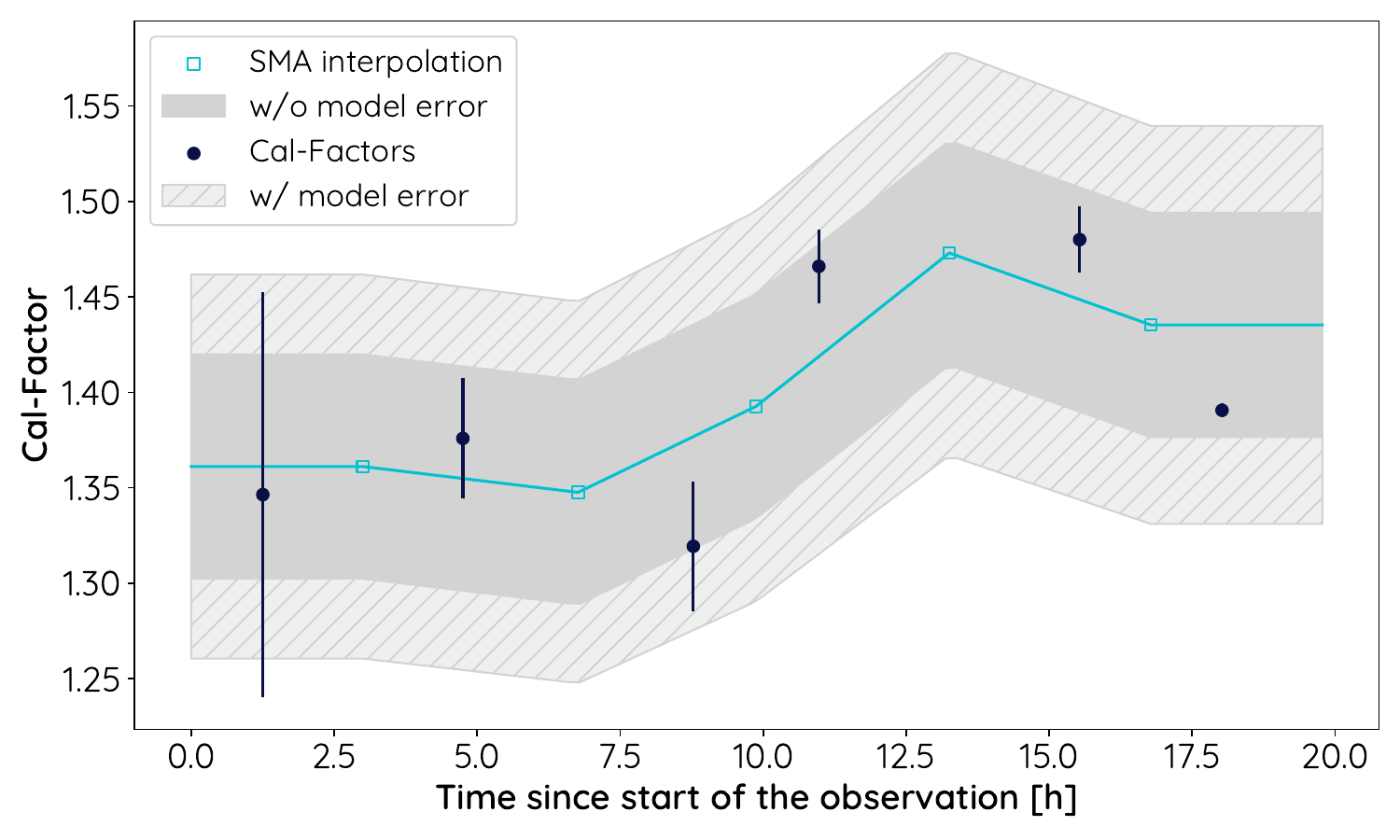}
    \caption{An example of the calibration factor evolution at 38.75\,GHz during the observing session on Oct 24, 2021. The black dots correspond to the (sub-)averaged calibration factor measurements $\Gamma_\mathrm{c,i}$. The blue line and blue squares indicate the simple moving average (SMA) interpolated calibration factor $\Gamma_\mathrm{SMA,i}$ (see Equation~\eqref{eq:smaInterpolation}), as used in the final analysis. The dark gray area represents the calibrator scatter $\sigma_{\mathrm{sc},\nu}$, while the light gray hashed area represents the total calibration uncertainty $\sigma_\mathrm{cal}$, including the model uncertainty $\sigma_\mathrm{model}$ and $\sigma_{\mathrm{sc},\nu}$.}
    \label{fig:calibration}
\end{figure}

As mentioned in Sect.~\ref{sec:observingprocess}, usually one of the secondary calibrator sources \review{is} observed every four hours during an observing session to determine the calibration factor. This calibration measurement typically consists of up to three scans per frequency on the calibrator. Usually, more than one scan of the calibrator is taken during a calibration measurement, therefore the mean and standard deviation are calculated to get a value with uncertainty for the calibration factor at the given time. In the case of only one calibrator measurement per calibration measurement the uncertainty is estimated depending on the frequency (see Sect.~\ref{sec:error}). For each observing session, one therefore gets a calibration factor every $\sim$4 hours, in the ideal case. 
An example is given in Fig.~\ref{fig:calibration}, where all calibration factor data points (black dots) are averages of the underlying (multiple) calibrator scans at the time. When talking about calibration factors in the following, it is referred to these, already (sub-)averaged calibration factors like the (black) ones presented in Fig.~\ref{fig:calibration}.

In principle, the calibration factor only depends on the sensitivity of the telescope at a given frequency and should therefore be constant throughout an observing session (for a given frequency). As one can see in Fig.~\ref{fig:calibration} this is not always the case. The calibration factor does change throughout the observing session, mostly due to temperature changes, which affect the ideal focus position. If the telescope is out of focus, the received signal is weaker and therefore the calibration factor rises. The telescope is being kept in focus by readjusting the focus every four hours (or more often in the case of significant temperature changes) to keep the calibration factor constant, but still the calibration factor varies throughout the observation. Modelling the behaviour of the focus is non-trivial, since it is unknown how much of the calibration factor fluctuation is statistical and how much is systematic due to, e.g., shifting focus. If the fluctuation was solely statistical, one would have to take the mean calibration factor throughout the entire session. If the fluctuation was solely systematic it would be best to linearly interpolate between the calibration factors. In the present case, however, interpolation between all data points would be an over-interpretation of the data and taking the mean cannot account for the fluctuation of the calibration factor. In order to take this fluctuation into account and also not to over-interpret the data points, the calibration factor is modelled by using a simple moving average (SMA) interpolation \citep{SMAInterpolation}. Essentially, this approach is a combination of both methods, since first, we take the mean between two adjacent calibration factor values and then interpolate between these mean values. If the calibration factor at time $t_\mathrm{i}$ is $\Gamma_\mathrm{c,i}$, the interpolated values are calculated via
\begin{equation}
\label{eq:smaInterpolation}
    \Gamma_\mathrm{SMA,i}=\frac{\Gamma_\mathrm{c,i}+\Gamma_\mathrm{c,i+1}}{2},
\end{equation}
where $\Gamma_\mathrm{c,i+1}$ is the calibration factor adjacent to $\Gamma_\mathrm{c,i}$. The same procedure is used for the time interpolation
\begin{equation}
    t_\mathrm{SMA,i}=\frac{t_\mathrm{i}+t_\mathrm{i+1}}{2}.
\end{equation}
If initially there are $n$ calibration factors $\Gamma_\mathrm{c,i}$, this interpolation will result in $n-1$ interpolated calibration factors $\Gamma_\mathrm{SMA,i}$. As indicated by the blue line in Fig.~\ref{fig:calibration}, we interpolate linearly between the $(t_\mathrm{SMA,i},\Gamma_\mathrm{SMA,i})$ values to get a general expression for the calibration factor at any given time. For times $t<t_{\mathrm{SMA},1}$, the calibration factor is modelled as constant $\Gamma_{\mathrm{SMA},1}$. Analogously, for times $t>t_{\mathrm{SMA},n-1}$, the calibration factor is modelled as constant $\Gamma_{\mathrm{SMA},n-1}$. If there is only one or two calibration factors available ($n=1,2$), a constant calibration factor is assumed for the entire epoch. For all other cases, the interpolation (blue line) is used to determine the calibration factor at any given time during the observing session. This means, for every source scan, one calculates the corresponding calibration factor $\Gamma_\mathrm{c}$ using the SMA-interpolation and then uses this calibration factor to calculate the flux density of the source $S_\mathrm{source}$ via
\begin{equation}
\label{eq:calibration}
    S_\mathrm{source}=T_\mathrm{src}\cdot\Gamma_\mathrm{c},
\end{equation}
where $T_\mathrm{src}$ is the observed source temperature.


\subsection{Discussion of uncertainties}
\label{sec:error}
In this section, we discuss the determination of the total flux density uncertainties $\sigma_\mathrm{tot}$. The final uncertainty has to include the main uncertainties due to the sub-scan fitting process with the data corrections $\sigma_\mathrm{fit}$, and the uncertainty of the calibration factor, $\sigma_\mathrm{cal}$.  

First, we focus on the uncertainty due to sub-scan fitting and data corrections, $\sigma_\mathrm{fit}$. Its value is calculated through Gaussian error propagation from the fitting and data correction process explained in Sect.~\ref{sec:pointcorr}. Note that the gain curve is assumed to be free of uncertainty here, since the accuracy of the gain curve is also reflected in the fluctuation of the calibration factors and therefore included in the calibration uncertainty $\sigma_\mathrm{cal}$. Usually, atmospheric corrections have the biggest impact on the data correction uncertainty, especially in sessions affected by bad weather. In total, $\sigma_\mathrm{fit}$ is on average on the order of $\sim 1$\,\%.

\begin{figure}
    \centering
    \includegraphics[width=\columnwidth]{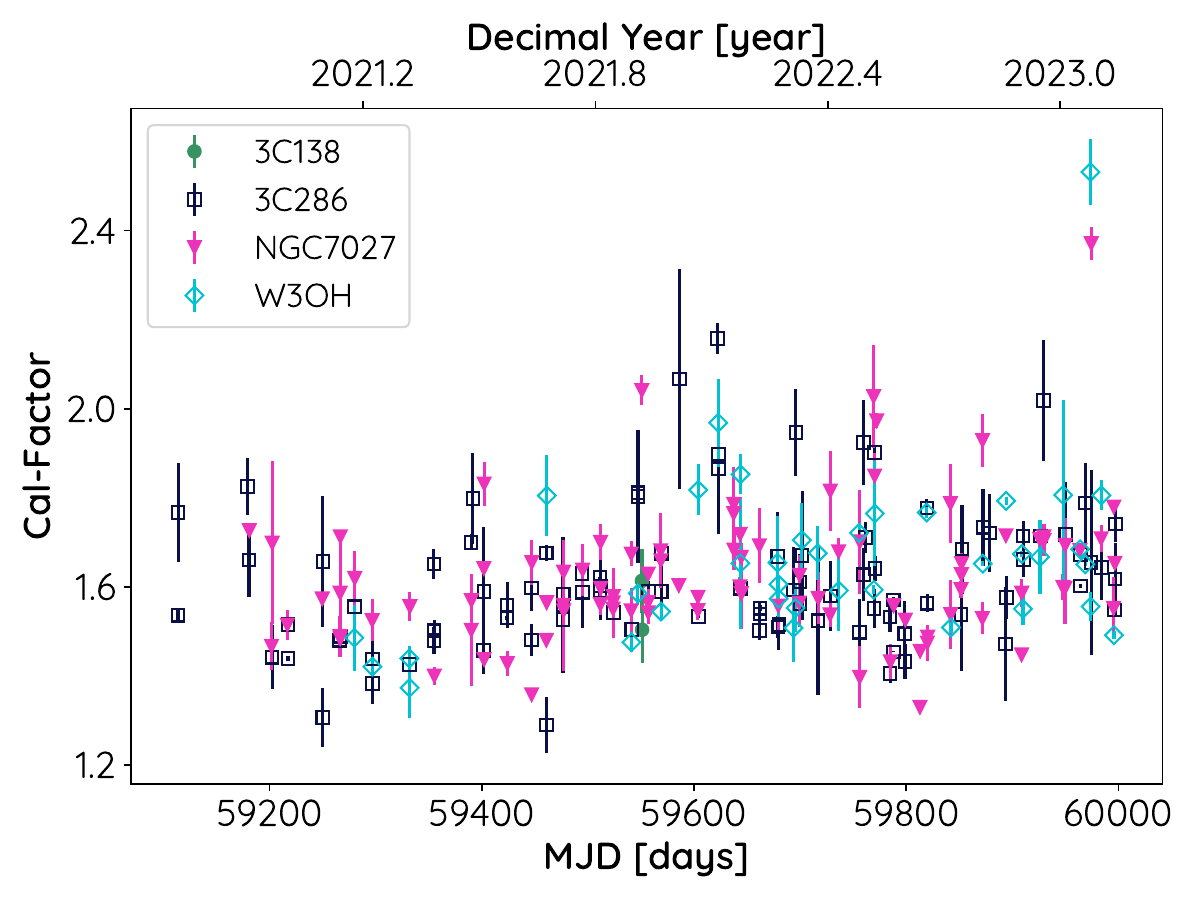}
    \caption{Illustrative calibration factor evolution at 41.25\,GHz throughout the entire program. We determine the calibration factor scattering uncertainty $\sigma_{\mathrm{sc},\nu}$ by taking the standard deviation of these values at each frequency. In this example, $\sigma_{\mathrm{sc},\nu}$ is on the order of 10\,\%. Significant outliers can be explained by bad weather epochs.}
    \label{fig:calibratorscatter}
\end{figure}

The main contribution of the flux density uncertainties comes from the calibration uncertainty, $\sigma_\mathrm{cal}$. There are two sources of uncertainty that determine the total calibration uncertainty. First, one needs to consider the fluctuation of the calibration factor $\sigma_{\mathrm{sc},\nu}$ for each frequency and observing epoch. Secondly, one needs to account for the uncertainty of the underlying calibrator model $\sigma_\mathrm{model}$. As explained in the previous section, an SMA-interpolation is used to calculate interpolated calibration factors. To be conservative, the uncertainty of the interpolated calibration factors is assumed to be constant. Therefore, the \review{root-mean-square} deviation (RMSD) of all (non-interpolated) calibration factors for each frequency and observing epoch is used as the uncertainty
\begin{equation}
    \sigma_{\mathrm{sc},\nu}=\sqrt{\frac{1}{n-1}\sum_{\mathrm{i}=1}^{n}(\Gamma_\mathrm{c,i}-\overline{\Gamma_\mathrm{c}})^2}.
\end{equation}
Here, $\overline{\Gamma_\mathrm{c}}$ is the mean over the non-interpolated calibration factors $\Gamma_\mathrm{c,i}$. This uncertainty is also illustrated in Fig.~\ref{fig:calibration} by the dark gray background. In the case where only one calibration factor was measured during an observing epoch, it is non-trivial to determine a sensible value for this uncertainty. In order to deal with this special case, the calibration factor scattering for each frequency from all observing epochs has been analyzed. An illustrative plot of the calibration factor evolution at 41.25\,GHz throughout the entire program since August 2020 is shown in Fig.~\ref{fig:calibratorscatter}. In order to derive a sensible uncertainty for these epochs, the standard deviation of all calibration factors throughout the program is used to get an estimate of the average calibration factor scatter. This analysis is performed for every frequency band. 
Following this procedure, a conservative calibration uncertainty of 5\,\% is used at 20\,mm and 14\,mm and 10\,\% at 7\,mm for epochs with only one calibration factor. In Fig.~\ref{fig:calibratorscatter}, one can also see that for some epochs the calibration factor varies more than for others. This is due to varying focus because of significant temperature and weather fluctuations (on shorter time scales than the usual focus adjustment interval every $\sim$4\,hours) in some observing sessions.

In addition to the calibration factor scattering $\sigma_{\mathrm{sc},\nu}$, one also needs to take into account the accuracy of the calibrator models $\sigma_\mathrm{model}$. The models of 3C\,295, 3C\,138 and 3C\,286 by \cite{Pearly2017} have an estimated accuracy of 3\,\% to 5\,\%, with the larger uncertainty at the lowest ($\sim50$\,MHz) and highest ($\sim50$\,GHz) ends. Since the observations take place from 14\,GHz$-$44\,GHz, which is at the higher end of their scale, an uncertainty of 5\,\% for the 3C\,295, 3C\,138 and 3C\,286 models is assumed. For the timely variable model of NGC\,7027, \cite{Zijlstra2008} provide an uncertainty of 6\,\%. It is estimated that the uncertainty for the W3(OH) model is also in the same range, since it has a similar underlying free-free emission model. As a conservative estimate, a general accuracy of the calibrator models of $\sigma_\mathrm{model}/\Gamma_\mathrm{c}=6$\,\% is assumed, which is the maximum uncertainty out of the models. One must not use the Gaussian law of error propagation to combine the uncertainties of the different models, since they are not statistically independent and all based on the same flux density scale by \cite{Baars1977}.
To calculate the total calibration uncertainty $\sigma_{\mathrm{cal}}$, the estimated model uncertainty $\sigma_{\mathrm{model}}$ and the calibration scatter $\sigma_{\mathrm{sc},\nu}$, which is individual for every epoch and frequency, are added quadratically. The total calibration uncertainty is therefore given by
\begin{equation}
    \sigma_\mathrm{cal}=\sqrt{\sigma_{\mathrm{model}}^2+\sigma_{\mathrm{sc},\nu}^2}.
\end{equation}
In order to obtain the total flux density uncertainty $\sigma_\mathrm{tot}$, one needs to combine $\sigma_\mathrm{cal}$ with the fitting uncertainty $\sigma_\mathrm{fit}$. This is done by Gaussian error propagation. Following Equation~\eqref{eq:calibration}, one finds
\begin{equation}
    \frac{\sigma_\mathrm{tot}}{S_\mathrm{source}}=\sqrt{\left(\frac{\sigma_\mathrm{cal}}{\Gamma_\mathrm{c}}\right)^2+\left(\frac{\sigma_\mathrm{fit}}{T_\mathrm{src}}\right)^2}
\end{equation}
for the total flux density uncertainty. Note that for the following analysis the model uncertainty is considered to be zero, since it is purely systematic throughout the entire program. Using it would therefore lead to an overestimation of the statistical flux density uncertainty. If the presented flux density values were to be combined with other radio data (using a different flux density scale), or if the absolute flux density values are of interest, the model uncertainty has to be taken into account.

\section{Results}
\label{sec:results}

\subsection{Median spectral indices}
In order to characterize the TELAMON TeV-sample in terms of the spectral index, we calculate spectral indices for every observation and present their medians in Table\,\ref{tab:average_table}. For some sources there is no sufficient frequency coverage available (i.e., less then three individual frequencies detected). Therefore, we could determine a sensible spectral index for 43 of our sample sources.

\begin{figure}
    \centering
    \includegraphics[width=\columnwidth]{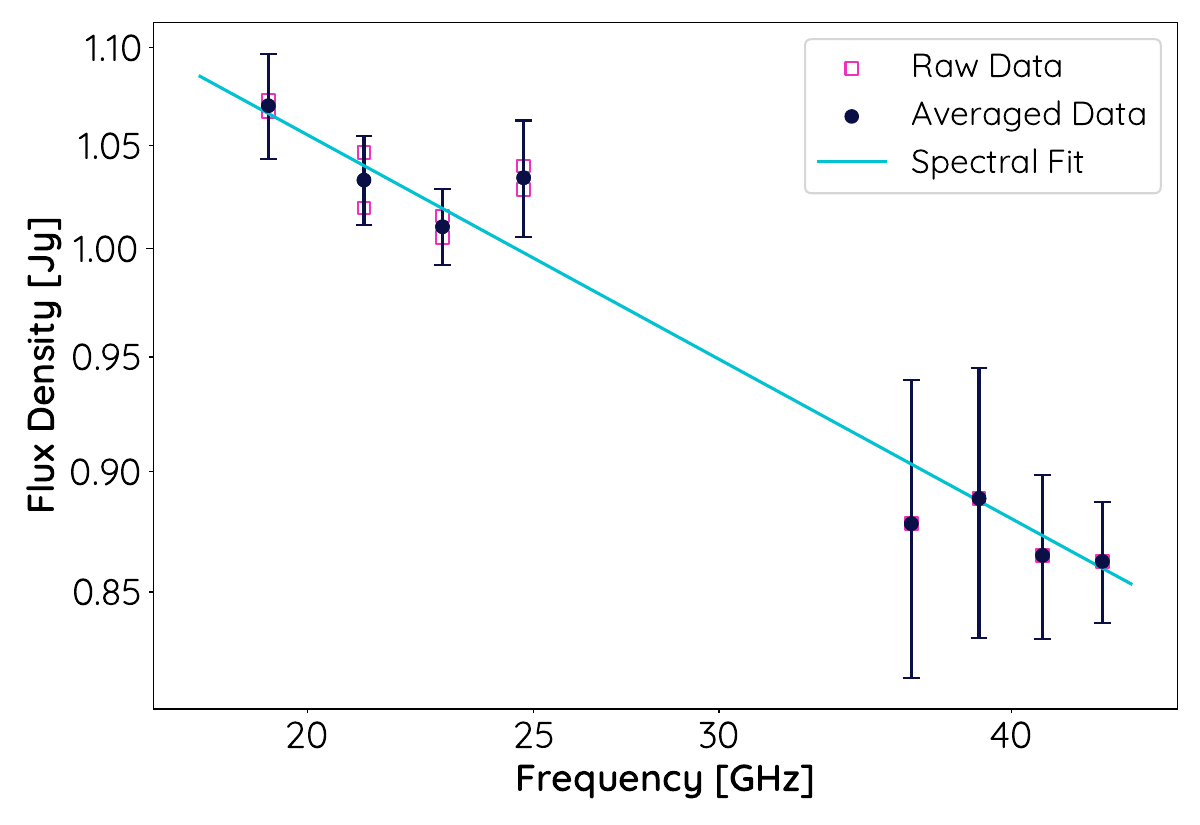}
    \caption{Example of the \reviewtwo{power-law} spectral fitting process (J1653+3945, \reviewtwo{also} known as Mrk\,501, on July 28, 2021). The magenta squares indicate the raw data, uncertainties are not shown to enhance readability. For every frequency, where more than one raw flux density value is available, the flux densities are averaged (black dots). The averaged flux densities are then fitted with a power-law spectrum (blue line). In the presented case, one finds $\alpha=-0.262\pm0.021$.}
    \label{fig:spectralfit}
\end{figure}

\subsubsection{Spectral index calculation}
\label{sec:spectralindexfit}

As established earlier, there can be two measurements of a source per epoch and frequency, which means that there exists more than one flux density value for the same frequency. This can be seen in the example presented in Fig.~\ref{fig:spectralfit} (magenta squares). In this case, the average flux density per frequency is calculated by taking the mean. If there is only one flux density value per frequency available, it is used as-is. The averaged values are depicted as black dots in Fig.~\ref{fig:spectralfit}. The uncertainty is determined by Gaussian error propagation, since distinct scans are considered independent measurements. After this first averaging process, a spectral power-law fit is performed to the data for every source at every observed epoch. This power-law is defined via
\begin{equation}
    S(\nu)\propto\nu^\alpha,
\label{eq:powerlaw}
\end{equation} 
where $S$ is the source flux density, $\nu$ the observed frequency and $\alpha$ the spectral index. The fit is performed using a Levenberg-Marquardt fitting algorithm \citep{LevenbergMarquardt} \review{which ensures a robust least-squares optimization. Moreover, it allows for bounds on the fitting parameters, which is important for the sub-band averaging (cf. Sec.\,\ref{sec:subbandaveraging})}. An example of such a fit for the source J1653+3945 (Mrk\,501) on July 28, 2021 is shown in Fig.~\ref{fig:spectralfit}, which results in a spectral index $\alpha=-0.262\pm0.021$. This analysis is applied to the entire data set, where at least three sub-band detections across all receivers (i.e., 14\,GHz$-$45\,GHz) are available for the same epoch and source. \review{\footnote{\review{In some cases (especially at 7\,mm) a source was occasionally observed but not detected (e.g., due to poor weather conditions). In principle, this non-detection yields information on an upper limit to the flux density of the source which could be used to put additional constraints on the spectral index. However, due to oftentimes rapidly changing weather conditions (especially cloud coverage) the determination of such upper limits is not reliably reproducible. We discuss the detectability of sources further in Sec.\,\ref{sec:discussion_detectrates} where we derive flux density thresholds for which one can expect a significant detection for each receiver. Using these thresholds as estimates for the upper flux-density limits in case of non-detections, it becomes clear that they are too high to have any significant impact on the spectral index fit.}}}

In order to get a typical value for the spectral index for every source, we list their median spectral indices in Table\,\ref{tab:average_table}, taken over all epochs. This is done only for sources for which spectral indices could be determined in three or more independent epochs to avoid outliers.

\subsubsection{Spectral indices in the TELAMON sample}

\begin{figure}
    \centering
    \includegraphics[width=\columnwidth]{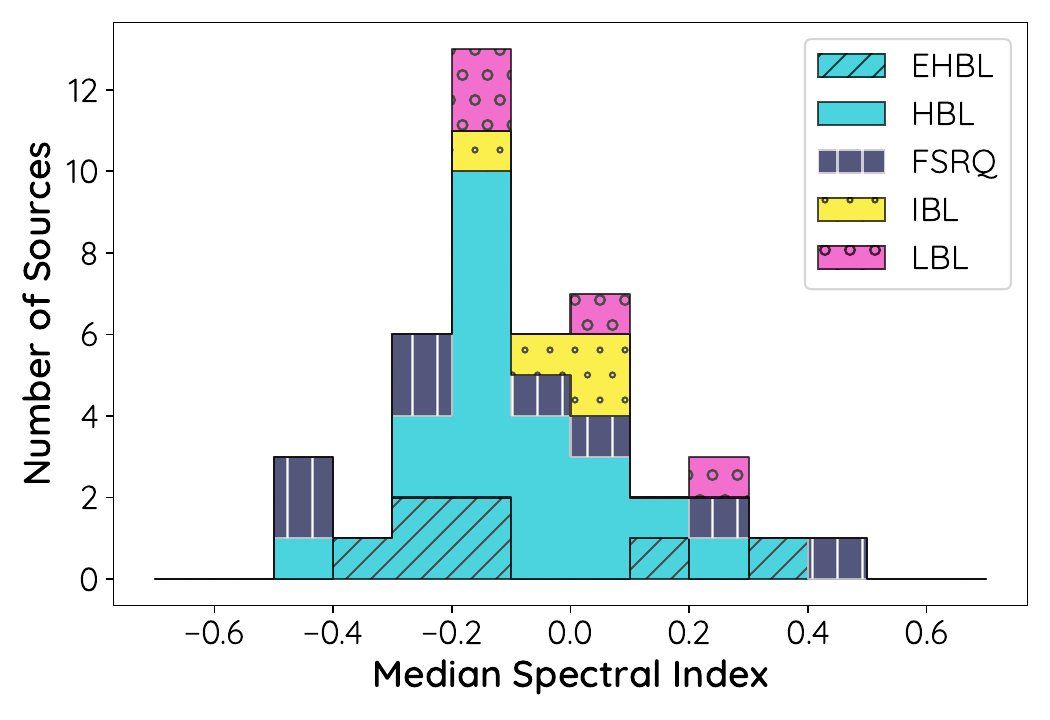}
    \caption{Distribution of the median spectral index across all available bands (i.e., 14\,GHz$-$45\,GHz), as found in the TELAMON sample sources. \reviewtwo{The histogram is divided into the source classes EHBL (blue with diagonal stripes), HBL (blue), IBL (yellow with dots), LBL (magenta with circles) and FSRQ (black with vertical stripes).}}
    \label{fig:spec_ind_hist}
\end{figure}

All median spectral indices across all bands (i.e., 14\,GHz$-$45\,GHz) are presented in Table\,\ref{tab:average_table}, and in Fig.\,\ref{fig:spec_ind_hist} as a histogram plot binned according to the source type. We find an overall average spectral index of $-$0.07, with a standard deviation of 0.20 and a median of $-$0.11. This is consistent with the expectation of a flat spectrum. We therefore conclude that the TeV-selected sub-sample of blazars presented here does not show any unexpected or special spectral features. In order to test if the spectral index distribution of HBL and EHBL is statistically different, we perform a two-sample Kolmogorov-Smirnov-test (KS-Test) which leads to a p-value of \review{$p=0.22$}. This  \review{indicates} that the distributions cannot be \review{clearly distinguished} and \review{are consistent with having} the same underlying statistics. For all other source classes, the sample size is not sufficient to perform significant statistical studies.

\begin{table*}
\caption{\small Average flux densities for all sources from the TELAMON sample as observed during the first 2.5 years. The median spectral index has been calculated across all receiver bands and epochs. Detection rates and the number of epochs each source was observed are shown.}
\label{tab:average_table}
\centering
\footnotesize
\resizebox{17.2cm}{!}{
\begin{threeparttable}
\begin{tabular}{@{}cc@{\,}ccccccccc@{~~}}
\hline\hline
ID  & $S_\textrm{20\,mm}$\tnote{a} & D-Rate\tnote{b} & N$_\mathrm{obs}$\tnote{c} & $S_\textrm{14\,mm}$\tnote{a} & D-Rate\tnote{b} & N$_\mathrm{obs}$\tnote{c} & $S_\textrm{7\,mm}$\tnote{a} & D-Rate\tnote{b} & N$_\mathrm{obs}$\tnote{c} & $\alpha$\tnote{d}\\
(J2000) & [Jy] & [\%] & & [Jy] & [\%] & & [Jy] & [\%] & &\\
\hline
J0035+5950 & 0.0725\,$\pm$\,0.0074 & 88 & 18 & 0.0676\,$\pm$\,0.0055 & 85 & 14 & - & 0 & 4 & $-$0.35 \\
J0112+2244 & - & - & 0 & 1.47\,$\pm$\,0.36 & 95 & 48 & 1.57\,$\pm$\,0.42 & 93 & 47 & 0.10 \\
J0136+3906 & 0.0413\,$\pm$\,0.0051 & 83 & 6 & 0.0397\,$\pm$\,0.0022 & 100 & 1 & - & - & 0 & $-$ \\
J0152+0146\review{\tnote{e}} & - & 0 & 3 & - & - & 0 & - & - & 0 & $-$ \\
J0214+5144 & 0.167\,$\pm$\,0.012 & 93 & 15 & 0.151\,$\pm$\,0.015 & 82 & 23 & 0.162\,$\pm$\,0.014 & 12 & 8 & $-$0.42 \\
J0221+3556 & - & - & 0 & 0.609\,$\pm$\,0.043 & 95 & 21 & 0.465\,$\pm$\,0.063 & 100 & 21 & $-$0.44 \\
J0222+4302 & - & - & 0 & 0.94\,$\pm$\,0.12 & 100 & 30 & 0.82\,$\pm$\,0.17 & 86 & 29 & $-$0.18 \\
J0232+2017 & 0.0413\,$\pm$\,0.0051 & 79 & 24 & 0.0377\,$\pm$\,0.0074 & 52 & 25 & - & 0 & 4 & 0.32 \\
J0316+4119 & 0.1386\,$\pm$\,0.0070 & 94 & 17 & 0.135\,$\pm$\,0.013 & 86 & 30 & - & 0 & 12 & $-$0.15 \\
J0319+1845 & 0.0202\,$\pm$\,0.0021 & 50 & 2 & - & 0 & 1 & - & - & 0 & $-$ \\
J0416+0105 & 0.0535\,$\pm$\,0.0081 & 85 & 20 & 0.051\,$\pm$\,0.010 & 68 & 19 & - & 0 & 6 & $-$0.16 \\
J0507+6737 & 0.0278\,$\pm$\,0.0013 & 25 & 4 & 0.0277\,$\pm$\,0.0016 & 22 & 9 & - & 0 & 2 & $-$ \\
J0509+0541 & - & - & 0 & 1.33\,$\pm$\,0.43 & 97 & 40 & 1.17\,$\pm$\,0.41 & 89 & 39 & $-$0.12 \\
J0521+2112 & - & - & 0 & 0.373\,$\pm$\,0.029 & 97 & 33 & 0.366\,$\pm$\,0.048 & 74 & 31 & 0.01 \\
J0648+1516 & 0.0324\,$\pm$\,0.0064 & 100 & 1 & 0.0312\,$\pm$\,0.0024 & 100 & 1 & - & - & 0 & $-$ \\
J0650+2502 & 0.0879\,$\pm$\,0.0082 & 77 & 18 & 0.0860\,$\pm$\,0.0099 & 85 & 21 & - & 0 & 4 & $-$0.13 \\
J0710+5909 & 0.05440\,$\pm$\,0.00042 & 20 & 10 & 0.0452\,$\pm$\,0.0022 & 100 & 1 & - & - & 0 & $-$ \\
J0721+7120 & - & - & 0 & 0.83\,$\pm$\,0.17 & 100 & 8 & 0.79\,$\pm$\,0.19 & 87 & 8 & $-$0.04 \\
J0733+5153$^\mathrm{\review{f}}$ & \review{-} & \review{0} & 9 & - & - & 0 & - & - & 0 & $-$ \\
J0739+0136 & - & - & 0 & 1.55\,$\pm$\,0.34 & 100 & 10 & 1.70\,$\pm$\,0.35 & 70 & 10 & 0.03 \\
J0809+5219 & 0.146\,$\pm$\,0.011 & 80 & 10 & 0.135\,$\pm$\,0.014 & 90 & 10 & - & 0 & 1 & $-$0.02 \\
J0812+0237 & 0.0403\,$\pm$\,0.0050 & 70 & 17 & 0.0438\,$\pm$\,0.0095 & 50 & 6 & - & - & 0 & $-$ \\
J0847+1133\review{\tnote{e}} & - & 0 & 2 & - & 0 & 1 & - & - & 0 & $-$ \\
J0854+2006 & - & - & 0 & 6.03\,$\pm$\,0.93 & 100 & 9 & 5.6\,$\pm$\,1.1 & 88 & 9 & $-$0.12 \\
J0958+6533 & - & - & 0 & 1.90\,$\pm$\,0.30 & 80 & 10 & 2.49\,$\pm$\,0.61 & 88 & 9 & 0.43 \\
J1015+4926 & 0.243\,$\pm$\,0.014 & 92 & 14 & 0.224\,$\pm$\,0.014 & 90 & 22 & 0.215\,$\pm$\,0.041 & 33 & 6 & $-$0.24 \\
J1058+2817 & 0.095\,$\pm$\,0.011 & 83 & 18 & 0.100\,$\pm$\,0.011 & 77 & 18 & - & 0 & 1 & 0.19 \\
J1104+3812 & 0.465\,$\pm$\,0.021 & 100 & 2 & 0.424\,$\pm$\,0.046 & 95 & 45 & 0.419\,$\pm$\,0.062 & 67 & 40 & $-$0.05 \\
J1136+7009 & 0.1628\,$\pm$\,0.0057 & 86 & 15 & 0.151\,$\pm$\,0.020 & 88 & 25 & - & 0 & 7 & $-$0.23 \\
J1136+6737 & 0.0350\,$\pm$\,0.0040 & 77 & 9 & - & 0 & 1 & - & - & 0 & $-$ \\
J1159+2914 & - & - & 0 & 4.5\,$\pm$\,2.1 & 100 & 10 & 4.9\,$\pm$\,2.2 & 80 & 10 & $-$0.07 \\
J1217+3007 & 0.374\,$\pm$\,0.024 & 100 & 2 & 0.404\,$\pm$\,0.050 & 96 & 32 & 0.410\,$\pm$\,0.059 & 89 & 29 & $-$0.05 \\
J1221+3010 & 0.0555\,$\pm$\,0.0064 & 93 & 16 & 0.065\,$\pm$\,0.016 & 75 & 12 & - & 0 & 2 & 0.28 \\
J1221+2813 & 0.521\,$\pm$\,0.052 & 50 & 2 & 0.490\,$\pm$\,0.056 & 94 & 39 & 0.458\,$\pm$\,0.065 & 75 & 36 & $-$0.10 \\
J1224+2122 & - & - & 0 & 0.89\,$\pm$\,0.15 & 90 & 10 & 0.624\,$\pm$\,0.078 & 88 & 9 & $-$0.49 \\
J1224+2436 & 0.0324\,$\pm$\,0.0087 & 90 & 10 & - & - & 0 & - & - & 0 & $-$ \\
J1230+2518 & 0.2639\,$\pm$\,0.0049 & 100 & 1 & 0.342\,$\pm$\,0.063 & 93 & 30 & 0.369\,$\pm$\,0.092 & 84 & 26 & 0.05 \\
J1415+1320 & 0.657\,$\pm$\,0.021 & 100 & 2 & 0.564\,$\pm$\,0.066 & 87 & 16 & 0.54\,$\pm$\,0.12 & 80 & 15 & $-$0.13 \\
J1422+3223 & 0.675\,$\pm$\,0.023 & 100 & 2 & 0.73\,$\pm$\,0.18 & 94 & 39 & 0.81\,$\pm$\,0.22 & 91 & 37 & 0.21 \\
J1427+2348 & 0.416\,$\pm$\,0.042 & 100 & 1 & 0.362\,$\pm$\,0.051 & 100 & 32 & 0.335\,$\pm$\,0.073 & 60 & 30 & $-$0.11 \\
J1428+4240 & 0.0304\,$\pm$\,0.0046 & 33 & 6 & 0.0225\,$\pm$\,0.0021 & 7 & 14 & - & 0 & 2 & $-$ \\
J1443+1200 & 0.0403\,$\pm$\,0.0035 & 77 & 9 & 0.0417\,$\pm$\,0.0060 & 100 & 1 & - & - & 0 & $-$ \\
J1443+2501 & 0.260\,$\pm$\,0.057 & 86 & 29 & 0.221\,$\pm$\,0.057 & 97 & 34 & - & 0 & 4 & $-$0.26 \\
J1555+1111 & 0.368\,$\pm$\,0.035 & 92 & 27 & 0.339\,$\pm$\,0.043 & 97 & 42 & 0.313\,$\pm$\,0.061 & 50 & 14 & $-$0.01 \\
J1653+3945 & 1.187\,$\pm$\,0.034 & 100 & 2 & 1.057\,$\pm$\,0.051 & 100 & 35 & 0.885\,$\pm$\,0.085 & 84 & 32 & $-$0.27 \\
J1725+1152 & 0.116\,$\pm$\,0.015 & 90 & 10 & 0.125\,$\pm$\,0.028 & 90 & 10 & - & - & 0 & 0.18 \\
J1728+5013 & 0.127\,$\pm$\,0.010 & 90 & 21 & 0.123\,$\pm$\,0.015 & 91 & 34 & - & 0 & 11 & $-$0.11 \\
J1743+1935 & 0.208\,$\pm$\,0.015 & 90 & 20 & 0.191\,$\pm$\,0.015 & 100 & 33 & - & 0 & 11 & $-$0.19 \\
J1751+0938 & 2.25\,$\pm$\,0.33 & 100 & 6 & 2.35\,$\pm$\,0.39 & 100 & 13 & 2.43\,$\pm$\,0.51 & 100 & 13 & 0.07 \\
J1813+3144 & 0.106\,$\pm$\,0.010 & 83 & 18 & 0.1070\,$\pm$\,0.0099 & 92 & 27 & - & 0 & 4 & $-$0.21 \\
J1943+2118 & 0.0377\,$\pm$\,0.0076 & 61 & 18 & 0.0316\,$\pm$\,0.0022 & 11 & 9 & - & - & 0 & $-$ \\
J1959+6508 & 0.204\,$\pm$\,0.019 & 90 & 21 & 0.197\,$\pm$\,0.017 & 86 & 36 & 0.2122\,$\pm$\,0.0091 & 18 & 11 & $-$0.15 \\
J2001+4352 & 0.223\,$\pm$\,0.056 & 100 & 14 & 0.230\,$\pm$\,0.053 & 100 & 14 & - & - & 0 & 0.05 \\
J2039+5219 & 0.0285\,$\pm$\,0.0039 & 100 & 1 & - & 0 & 1 & - & - & 0 & $-$ \\
J2056+4940 & 0.117\,$\pm$\,0.018 & 77 & 9 & 0.111\,$\pm$\,0.017 & 66 & 9 & - & - & 0 & $-$0.13 \\
J2202+4216 & - & - & 0 & 5.8\,$\pm$\,2.3 & 90 & 11 & 7.1\,$\pm$\,2.7 & 90 & 11 & 0.27 \\
J2250+3825\review{\tnote{e}} & - & 0 & 2 & - & - & 0 & - & - & 0 & $-$ \\
J2243+2021 & 0.0987\,$\pm$\,0.0058 & 70 & 20 & 0.0988\,$\pm$\,0.0087 & 78 & 19 & - & - & 0 & 0.04 \\
J2347+5142 & 0.1615\,$\pm$\,0.0097 & 90 & 20 & 0.151\,$\pm$\,0.014 & 87 & 33 & - & 0 & 12 & $-$0.29 \\
\hline
\end{tabular}
\begin{tablenotes}\footnotesize
    \tnote{a}Average flux density and standard deviation at the given wavelength taken over all observed epochs.
    \tnote{b}Detection rate (i.e., number of epochs with detection divided by the number of epochs where the source was observed) at the given wavelength.
    \tnote{c}Number of epochs where the source was observed at the given wavelength.
    \tnote{d}Median spectral index across all observed epochs with at least three sub-band detections.
    \tnote{e}Source was observed but not detected at 20\,mm and 14\,mm. We have confirmed with independent 45\,mm observations that the source is indeed very faint, i.e., a successful detection at wavelengths $<20$\,mm is unlikely. For these faint sources we have defined a new observing category and are now monitoring them with high detection rates at 45mm.
    \review{\tnote{f} Affected by source confusion with a brighter neighbouring source, therefore no detection was possible.}
\end{tablenotes}
\end{threeparttable}
}
\end{table*}

\subsection{Average flux densities}
In order to simplify the analysis, the observed flux densities of every source and epoch are averaged over the sub-frequencies of the 20\,mm, 14\,mm and 7\,mm receivers, respectively. This means for every source we derive an average flux density value at 20\,mm, 14\,mm, and 7\,mm for each epoch, given that the source was observed and detected in these bands. The main reason for this is that in some cases not all of the four (two for 20\,mm) sub-bands of each receiver show a significant source detection. This can be due to RFI, because the source is too weak at the highest frequencies, or due to background noise. Averaging over the sub-bands of each receiver will make it possible to compare all epochs with each other, even if a sub-band flux density value might be vacant in one or more epochs. However, taking the mean or the weighted mean of all measured values does not suffice to ensure the comparability of epochs. For example, if a source was detected at 19.25\,GHz and 21.15\,GHz at Epoch A and at another Epoch B at 22.85\,GHz and 24.75\,GHz, taking the mean will shift the mean frequency of this average. Since the spectrum of the sample sources is not always flat ($S(\nu)\neq$\,const.), this could turn out to be problematic when comparing flux densities from two distinct epochs. We therefore introduce a new method of obtaining average flux densities from the receiver sub-bands using spectral fits in the following section.

\subsubsection{Sub-band averaging}
\label{sec:subbandaveraging}

\begin{figure}
    \centering
    \includegraphics[width=\columnwidth]{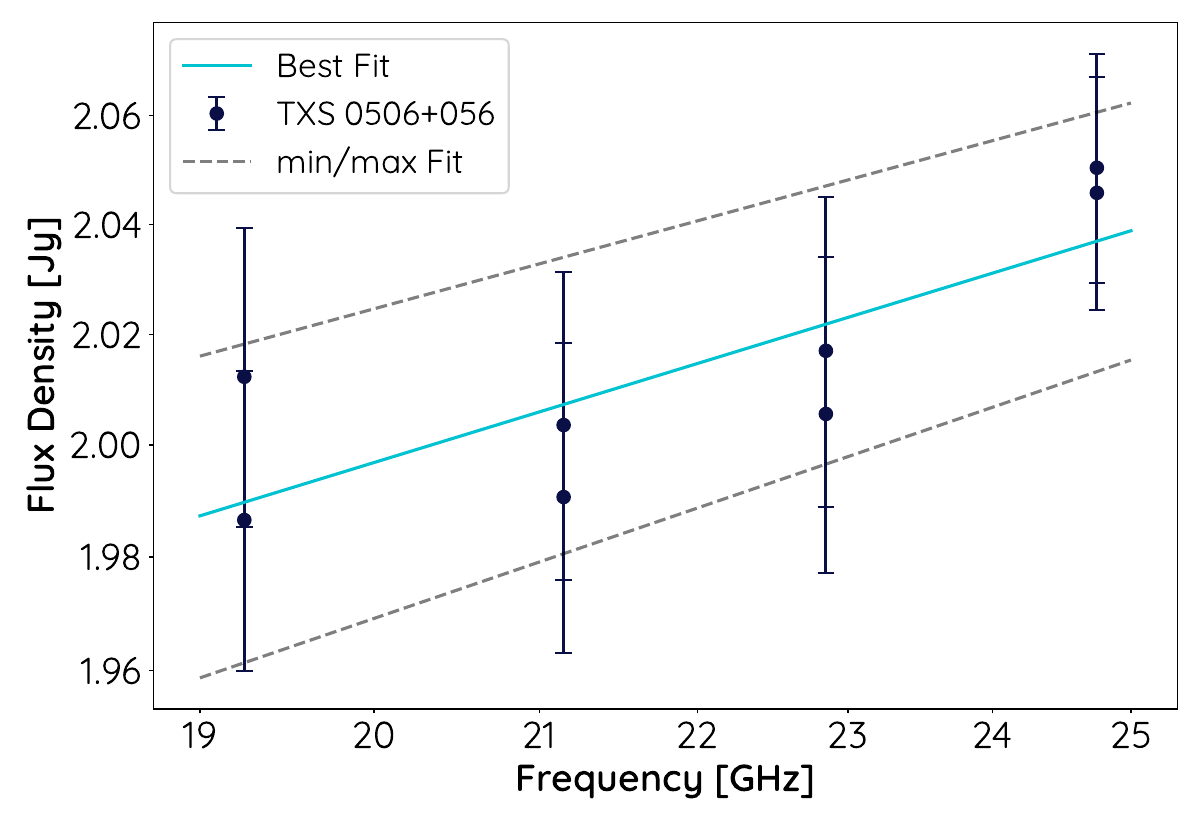}
    \caption{Example of the sub-band averaging process (J0509+0541, known as TXS\,0506+056 on Jan 2, 2021 \reviewtwo{with two distinct scans} at 14\,mm). The average flux density is calculated by taking the integral of the best fit between 19\,GHz and 25\,GHz. The uncertainties are determined by integrating the min/max fits.}
    \label{fig:average_fit_plot}
\end{figure}

For each receiver (7\,mm, 14\,mm and 20\,mm), at first a power-law spectral fit (cf. Eq.\,\eqref{eq:powerlaw}) is performed to the raw data within the receiver band width. Similar to the spectral index calculation in Sect.~\ref{sec:spectralindexfit}, a Levenberg-Marquart fitting algorithm is used to calculate the fit. Note that here, we set a bound of $|\alpha|<0.5$ to the spectral index, since the analysis presented here is more sensitive to outliers. From the fit, the average flux density $\overline{S}$ in the range of the receiver bandwidth is calculated:
\begin{equation}
\label{eq:average}
    \overline{S}=\frac{\int_{\nu_1}^{\nu_2}S(\nu)\textup{d}\nu}{\nu_2-\nu_1}.
\end{equation}
For the 20\,mm receiver, the integration limits are $\nu_1=14\,$GHz and $\nu_2=17\,$GHz, for the 14\,mm receiver $\nu_1=19\,$GHz and $\nu_2=25$\,GHz, and for the 7\,mm receiver $\nu_1=36$\,GHz and $\nu_2=44$\,GHz. This method ensures that the average flux density is always calculated from the same frequency range, with best knowledge about the intrinsic spectrum of the source for each epoch. In the case where \review{only} data for one sub-frequency (e.g., only for 19.25\,GHz) are available, \review{we use the data as-is. This is equivalent to assuming a flat spectrum ($\alpha=0$) over the receiver bandwidth, which} is typical for \review{compact radio jets} \citep[e.g.,][]{Zensus1997}. 

In order to define an uncertainty for the average flux density values, different approaches are used according to how many sub-bands are detected. In the case of two, three or four (i.e., all) sub-frequencies detected, two additional (uncertainty-)fits are performed: One with the uncertainties subtracted from the measured flux density values and the other one with the uncertainties added to the measured values (min- and max-fit). This is justified since in this case, the uncertainty mostly consists of systematic calibration uncertainty, i.e., the measured values may all together be higher (or lower) than the best values but relative to each other, they are known much more precisely. This procedure is illustrated in Fig.~\ref{fig:average_fit_plot}. The uncertainty is determined by the difference between the best fit to the minimum and maximum flux density (min/max-fit), and the best fit to the flux density value. In the case of only one sub-frequency detected, this is vastly different, since no sensible fit can be performed. 
The uncertainty of this value is estimated in a similar manner to the min/max-fits mentioned earlier. Again, the measurement uncertainties are subtracted/added to the flux density value. Then, different spectral distributions with $\alpha \in \left\{-0.5; 0 ; 0.5\right\}$, originating at the min/max flux densities, are considered. This means, one gets three alternate spectral distributions for minimum and maximum flux density, respectively. Analogous to the previous calculations, the difference between the integrated alternate spectral distributions and the integrated best fit (in this case the flat spectrum) is calculated. The maximum difference is then used as the average flux density uncertainty.

In the following sections, when referring to flux densities at 7\,mm, 14\,mm or 20\,mm, we always refer to the sub-band-averaged flux density in each band, calculated using the method introduced in this section, if not declared otherwise.

\subsubsection{Flux densities in the TELAMON sample}

\begin{figure}
    \centering
    \includegraphics[width=\columnwidth]{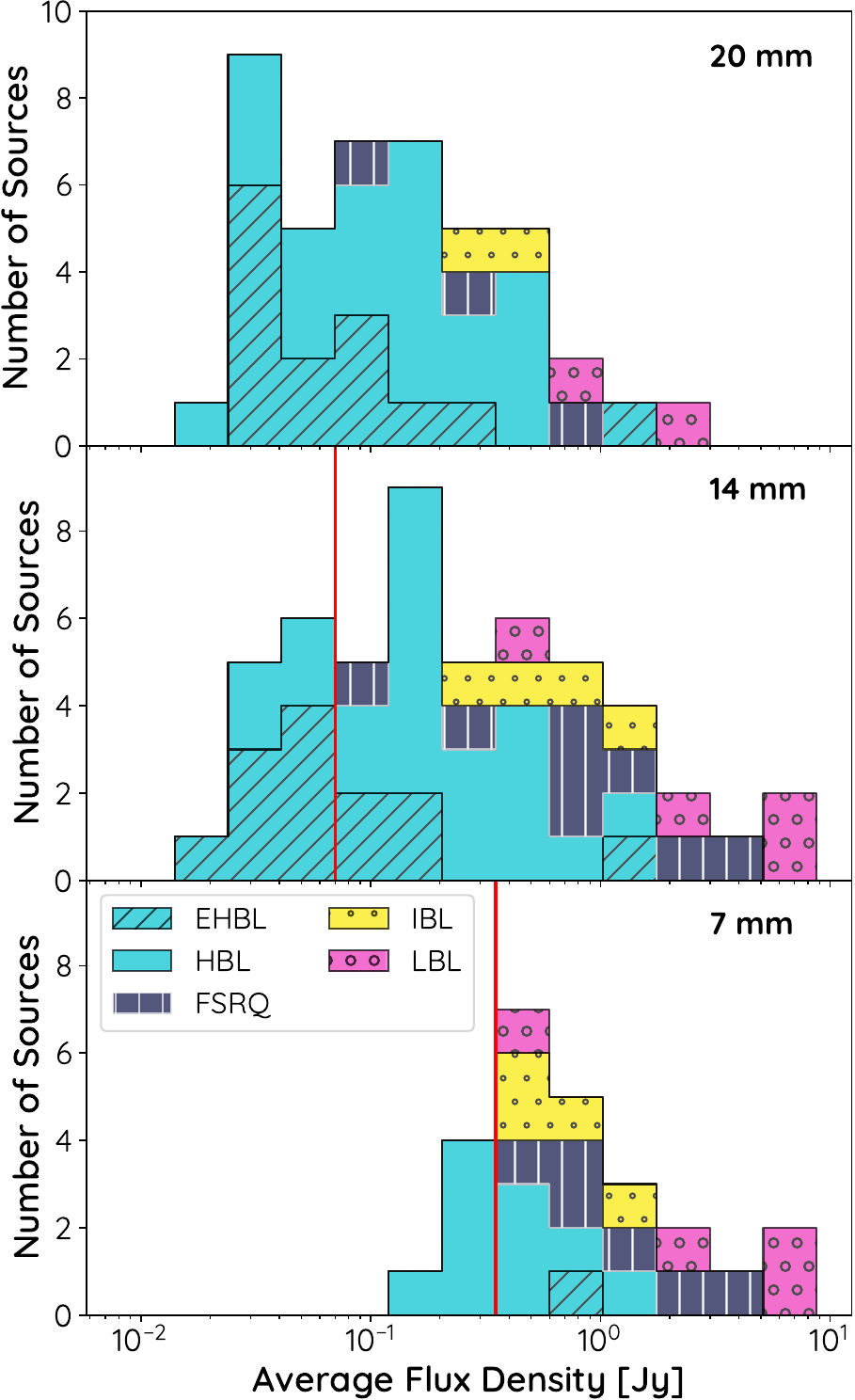}
    \caption{Distribution of the average flux density found in the TELAMON sample sources for the 20\,mm (top), 14\,mm (center), and 7\,mm (bottom) receivers. The histogram is divided into the source classes EHBL (blue with diagonal stripes), HBL (blue), IBL (yellow with dots), LBL (magenta with circles) and FSRQ (black with vertical stripes). The red vertical lines indicate the flux density above which one can expect a $>50$\,\% detection chance per epoch (see Sect.\,\ref{sec:discussion_detectrates}). Below these thresholds the statistics are significantly affected due to the limited telescope sensitivity. For 20\,mm we do not have enough data on the faintest sources to determine this limit.}
    \label{fig:averhist}
\end{figure}

The sub-band averaging procedure introduced in the previous section provides us with one flux density value at 20\,mm, 14\,mm and 7\,mm for every source at every epoch, given that it was observed and detected. In order to get an overview of the general source properties, we calculate the average flux densities of every source for 20\,mm, 14\,mm and 7\,mm over time from the sub-band averaged flux densities. This allows us to characterize the observed sample in terms of average source flux density. All average flux density values are presented in Table\,\ref{tab:average_table}. Their uncertainties correspond to the standard deviation of the sub-band averaged flux densities and therefore also reflect the intrinsic source variability. In the case where a source has only been detected once, the presented uncertainty is equal to the uncertainty of the sub-band averaged flux density. All average flux densities presented in Table\,\ref{tab:average_table} are also depicted as histogram plots in Fig.\,\ref{fig:averhist}, binned according to the source type. Here, one can see that the 7\,mm flux densities (Fig.\,\ref{fig:averhist}, bottom) are by far the smallest sample and limited to flux densities $\gtrsim$150\,mJy. Only 25 out of 41 observed sources show a significant detection at 7\,mm. This can be explained by the detection limit of the 7\,mm receiver (see Sect.\ref{sec:discussion_detectrates}) and by the fact that most of the sources in our sample are too faint to be detected at 7\,mm. This is vastly different at 20\,mm and 14\,mm. At 14\,mm (Fig.\,\ref{fig:averhist}, center), 51 of the 55 observed sources show a significant detection in at least one epoch. At 20\,mm (Fig.\,\ref{fig:averhist}, top), 44 of the 47 observed sources show a significant detection in at least one epoch. Only three observed sources (J0152+0146, J0847+1133, J2250+3825) have not been detected at all. We have confirmed with independent 45\,mm observations that these sources are amongst the faintest in our sample and therefore most likely below the detection limit of the telescope at shorter wavelengths with the current setup.

Following from the values presented in Table\,\ref{tab:average_table}, we find an average 7\,mm flux density of 1.4\,Jy with a standard deviation of 1.8\,Jy and a median of 0.6\,Jy in our sample. This can only be considered as an upper limit to the average 7\,mm flux density of the entire sample, since it is heavily biased by the non-detection of sources with flux densities $\lesssim$150\,mJy due to the sensitivity limit at 7\,mm. At 14\,mm, the average source flux density is $0.7$\,Jy with a standard deviation of $1.3$\,Jy and a median of $0.2$\,Jy. At 20\,mm, we find an average source flux density of $0.24$\,Jy with a standard deviation of $0.3\review{9}$\,Jy and a median of $0.12$\,Jy. At the latter two wavelengths, the receiver sensitivity also limits the detection of very faint sources, but since almost all sources from our sample were detected in these bands, we consider this bias to be much less significant than for 7\,mm.
This allows us to perform a sensible statistical comparison of the 20\,mm and 14\,mm data. For both wavelengths, we find that according to a two sample KS-test one cannot reject the null-hypothesis that EHBL and HBL have the same underlying statistical distribution (14\,mm: \review{$p=0.02$; 20\,mm: $p=0.19$}). However, the sub-sample of EHBLs seems to populate lower flux densities than the sub-sample of HBLs according to the median average flux densities\review{:} At 20\,mm, the subsample of EHBLs exhibits a median average flux density of \review{0.048\,Jy (mean: 0.15\,Jy, RMSD: 0.30\,Jy)}, while for HBLs one finds a median of \review{0.13\,Jy (mean: 0.16\,Jy, RMSD: 0.13\,Jy)}. At 14\,mm, the subsample of EHBLs exhibits a median average flux density of \review{0.045\,Jy (mean: 0.15\,Jy, RMSD: 0.28\,Jy)}, while for HBLs one finds a median of \review{0.15\,Jy (mean: 0.27\,Jy, RMSD: 0.31\,Jy)}. 
\review{The larger mean values are driven by one single EHBL (Mrk\,501) with unusually high flux density.
Considering instead the median flux densities, the impact of this single source is reduced and it appears that a majority of sources from the sub-sample of HBLs shows higher flux densities than the sub-sample of EHBLs, at moderate significance, which is in line with visual inspection of Fig.\,\ref{fig:averhist}.} This is consistent with the per definition higher synchrotron peak frequencies of EHBLs which \review{results in a shift of their SED towards higher frequencies. Consequently, we find that the high-synchrotron peaked sources are fainter in the radio band than the lower-peaked sources in our sample.}

\subsection{Detection rates}

In order to analyze the impact of the receiver sensitivity on the detectable source flux density, we calculate detection rates for all observed sources and receivers. For every source, we count the observing epochs (dates) during which the source was observed and the epochs where the source was detected in at least one sub-frequency of the given receiver. The detection rate is then given by the number of detections divided by the number of observations. The detection rates  and number of observations for each receiver and source are presented in Table\,\ref{tab:average_table} next to the average flux density values. These detection rates are further discussed in section \ref{sec:discussion_detectrates}.

\begin{figure*}
  \begin{minipage}[c]{0.67\textwidth}
    \centering
    \includegraphics[width=\textwidth]{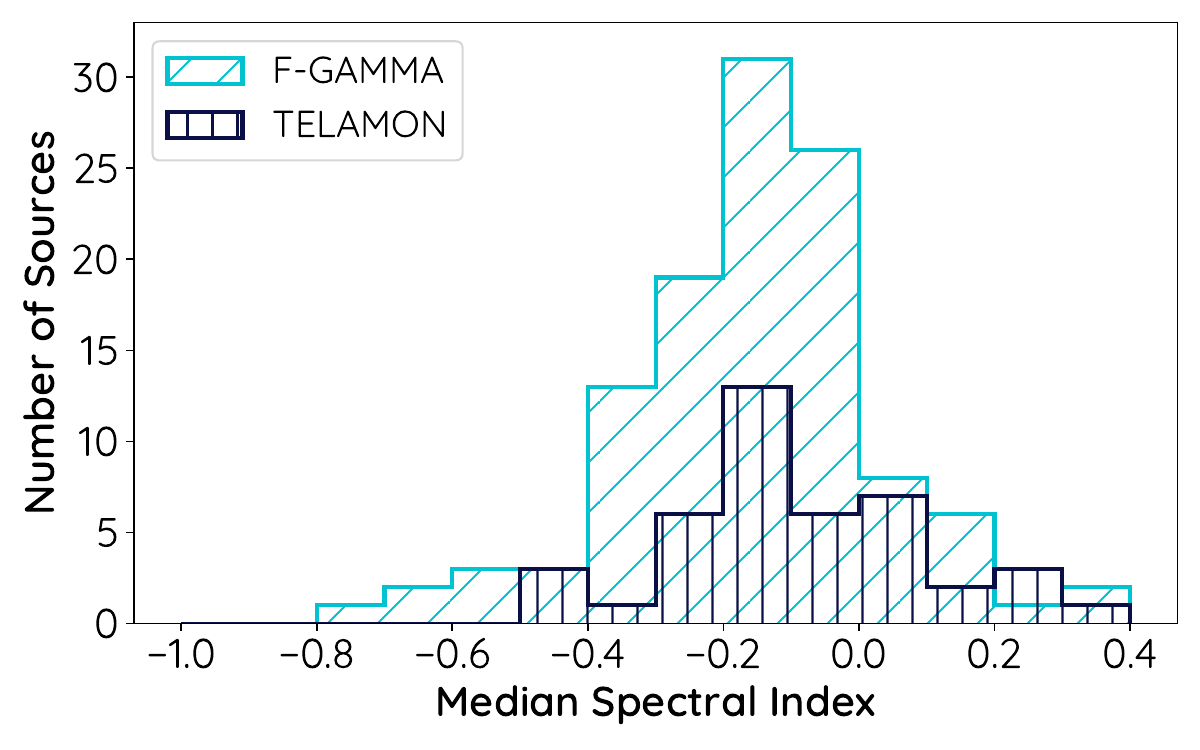}
  \end{minipage}\hfill
  \begin{minipage}[c]{0.3\textwidth}
    \caption{Histogram plot of median \reviewtwo{radio} spectral indices for all sources from the F-GAMMA sample \citep[][blue with diagonal stripes]{Angelakis2019} and the TELAMON (black with vertical stripes) sample between 14\,GHz and 45\,GHz.}
    \label{fig:TELAMON_FGAMMA_spec}
  \end{minipage}
\end{figure*}

\begin{figure*}
  \begin{minipage}[c]{0.67\textwidth}
  \centering
    \includegraphics[width=\textwidth]{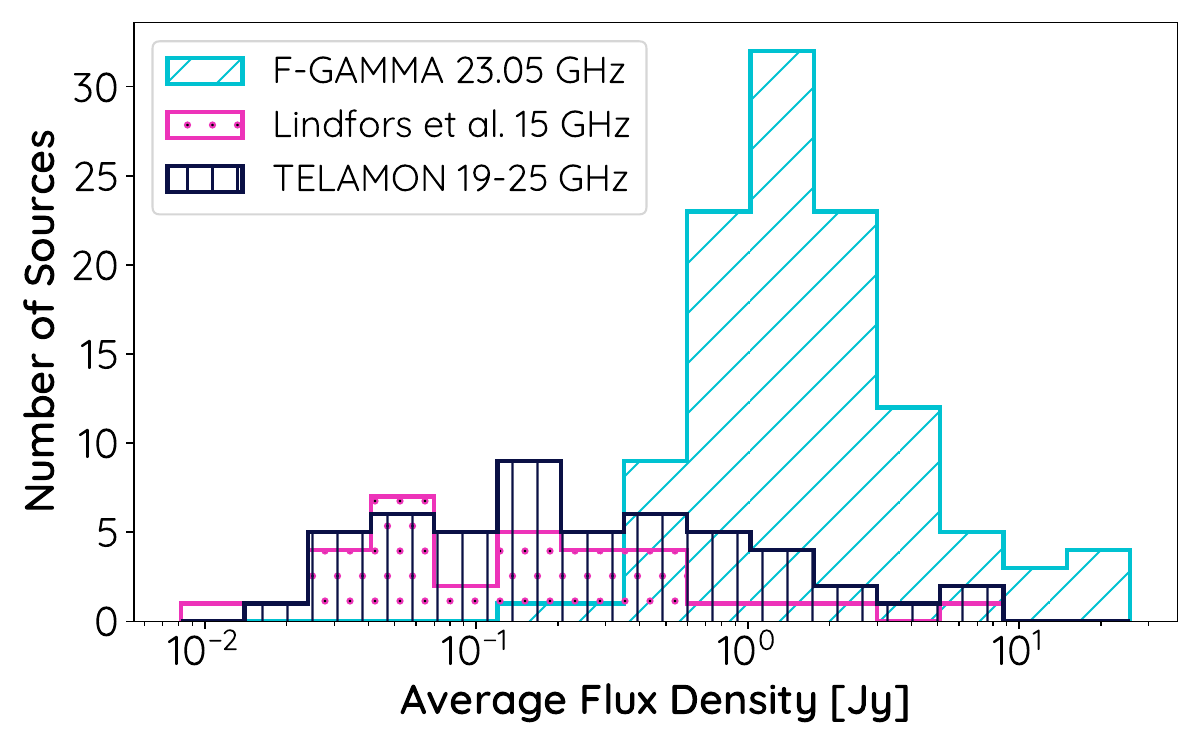}
  \end{minipage}\hfill
  \begin{minipage}[c]{0.3\textwidth}
    \caption{Comparison of the GeV-selected F-GAMMA \citep[][blue with diagonal stripes]{Angelakis2019} and the TeV-selected \cite{Lindfors} and TELAMON samples (magenta with dots and black with vertical stripes, respectively) according to the average source \reviewtwo{radio} flux density (F-GAMMA: 23.05\,GHz, Lindfors et al.: 15\,GHz, TELAMON: 14\,mm, i.e., 19\,GHz$-$25\,GHz).}
    \label{fig:TELAMON_FGAMMA_LINDFORS}
  \end{minipage}
\end{figure*}

\section{Discussion}
\label{sec:discussion}
In this section, we discuss and compare the TELAMON source properties with previous programs, namely the F-GAMMA monitoring \citep{Fuhrmann,FGAMMA_VariabilityDoppler,Angelakis2019} and a study on TeV-selected blazars by \cite{Lindfors}.

\subsection{Spectral index discussion}
Since the study by \cite{Lindfors} was carried out only at a single radio frequency (15\,GHz), we can only compare spectral indices with the F-GAMMA program. \cite{Angelakis2019} provide spectral indices for various frequency windows. \review{Their high-band (i.e., 14.6\,GHz$-$43\,GHz) is compatible with the TELAMON frequency range (i.e., 15\,GHz$-$44\,GHz). We therefore use the F-GAMMA high-band spectral indices provided by \cite{Angelakis2019} to compare the spectral indices of the TELAMON and F-GAMMA samples.} Note that the F-GAMMA data \citep{Angelakis2019} includes several (non-blazar) sources with a steep spectrum ($\alpha\lesssim-0.8$), which the TELAMON sample does not show. These are mostly calibrator sources which we exclude for further discussion. Figure\,\ref{fig:TELAMON_FGAMMA_spec} shows a direct comparison of the distribution of both samples in form of a histogram plot. The overall sample size of F-GAMMA (\review{118} sources with spectral index) is superior to the TELAMON sample (43 sources with spectral index). According to a two-sample KS-test, we find that the samples most likely have the same underlying distributions (\review{$p=0.24$}). This suggests that GeV-emitting blazars and TeV-emitting blazars exhibit similar spectral indices.

\subsection{Flux density discussion}

Similar to the spectral index comparison, we have calculated average flux densities for the F-GAMMA data at 23.05\,GHz from \cite{Angelakis2019} to compare with the TELAMON average flux densities at the overlapping wavelength of 14\,mm. On top of that, we use the average flux density values provided by \cite{Lindfors} at 15\,GHz for an additional comparison. All three flux density distributions are depicted in Fig.\,\ref{fig:TELAMON_FGAMMA_LINDFORS}. One can clearly see that the flux densities of the GeV-selected F-GAMMA sources are in general much higher than for the TeV-selected \cite{Lindfors} and TELAMON samples. This is consistent with the fact that most TeV emitting blazars have their SED shifted to higher energies and are consequently fainter in the radio band. We further performed a two-sample KS-test to compare TELAMON flux densities with the F-GAMMA and \cite{Lindfors} sample and find with high significance that the TELAMON and \cite{Lindfors} sources show a similar distribution ($p=0.30$) while we have to reject the null-hypothesis of similar underlying distributions for TELAMON and F-GAMMA ($p=1.4\times10^{-16}$). We therefore conclude that this difference in flux density seems to be characteristic for the target selection of either GeV- (F-GAMMA) or TeV-selected \citep[TELAMON,][]{Lindfors} blazars. TeV-emitting blazars seem to be fainter radio emitters than GeV-emitting blazars. Note that all three programs are limited by similar telescope sensitivity, which means we expect that this has no impact on the statistical comparison of the samples.

\subsection{Long-term monitoring strategy}

\label{sec:discussion_detectrates}
In order to address the question of the telescope sensitivity limit and to suggest a TeV-blazar sample well suited for long-term monitoring observations with the Effelsberg 100-m telescope, we discuss the relationship between source flux density and detection rate for all three receivers (20\,mm, 14\,mm and 7\,mm) in this section. One can see in Table\,\ref{tab:average_table} that for some very faint sources the detection rate is rather low (i.e., $\sim$10\,\%). Observing these sources in a long-term monitoring study would lead to a waste of telescope time that could instead be used to include other (brighter) sources with better detection potential more frequently. On top of that, in order to perform variability and multi-wavelength correlation studies, a sufficient sample size (flux density measurements per source) is required, which would take decades for sources with low detection rates and an observing cadence of two to four weeks. Moreover, such a sensitivity study can be useful to derive upper flux density limits in case of a non-detection in future studies. 

\begin{figure}
    \centering
    \includegraphics[width=\columnwidth]{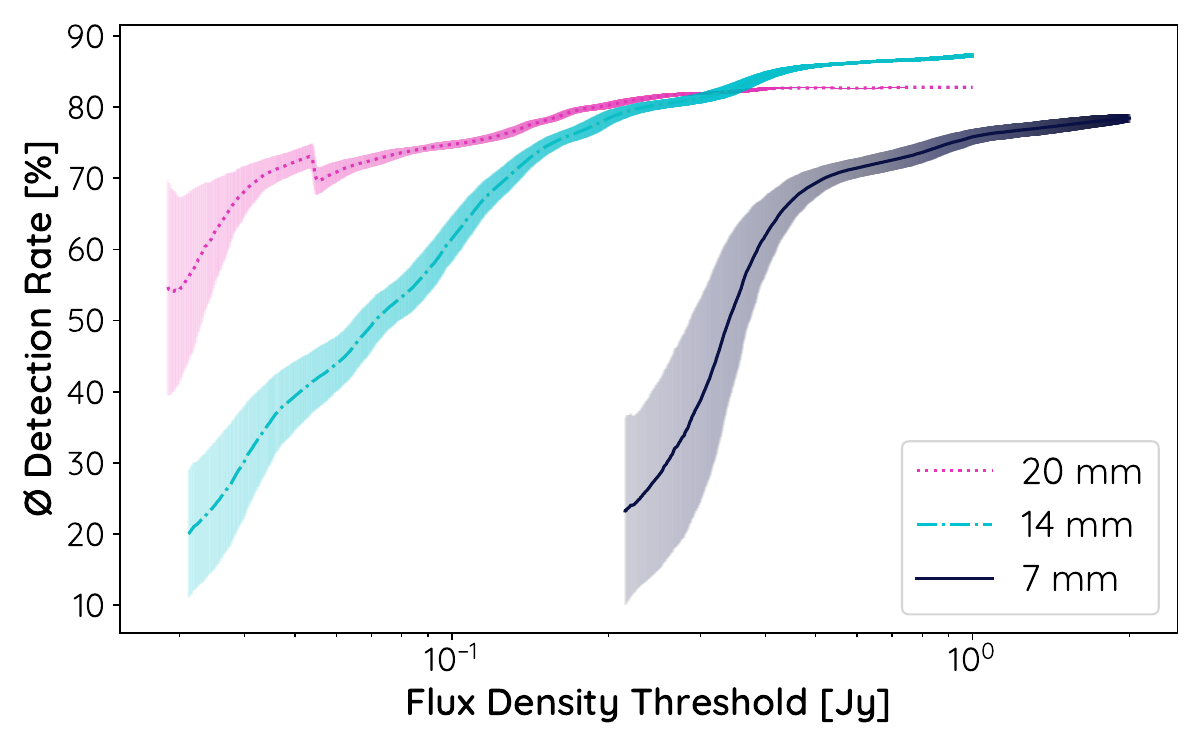}
    \caption{Average detection rate for sources below a given flux density threshold for the 20\,mm, 14\,mm and 7\,mm receiver as calculated from the first 2.5 years of TELAMON observations. The errors are calculated in a Monte-Carlo way, which takes into account the variability of the observed sources.}
    \label{fig:detectionrates}
\end{figure}

In principle, the detection rate should only depend on the sensitivity of the receiver and can be calculated by telescope intrinsic parameters. However, given that we are performing a long-term monitoring program, bad weather epochs and intrinsic source variability can influence the detectability of fainter sources, especially at the highest frequencies. To maximize source detections, we try to prioritize the faintest sources during good weather conditions, but this is not always possible since we are trying to keep a cadence of two to four weeks for all sources. In principle, the detection chances for the faintest sources can be increased by using more scans, i.e., spending more observing time per source. However, due to the limited observing time and large number of sources this was usually not feasible. We consider the 2.5\,year observations presented in this paper to be a good representation of such effects and therefore well suited to identify a flux density limit above which it is sensible to monitor sources as part of a long-term study with the given telescope and observing setup. For each receiver, we calculate the average detection rate for all sources below a certain flux density threshold. This is done for 10,000 different flux density thresholds. Moreover, this entire procedure is carried out 1,000 times with varying average source flux densities in a Monte-Carlo (MC) way, assuming the average flux densities and their uncertainties in Table\,\ref{tab:average_table} correspond to Gaussian distributions. The mean average detection rate and its standard deviation taken from the 1,000 MC-iterations is then plotted in Fig.\,\ref{fig:detectionrates} for 10,000 different flux density threshold values. In order to guarantee statistical convergence, we require a minimum of three sources to be included in the average detection rate count, therefore the average detection rates in Fig.\,\ref{fig:detectionrates} can only be presented above a certain flux density threshold. If we assume that the detection rate is monotonically increasing with flux density, the detection rate curves in Fig.\,\ref{fig:detectionrates} correspond to lower limits of the actual detection rate at any given flux density. For a sensible long term monitoring, we require a minimum detection rate of $50$\,\% (i.e., a source detection at least every other epoch) to optimize the scientific outcome of the observing time and to prevent too many non-detections. Using the data from Fig.\,\ref{fig:detectionrates}, we can determine a flux density limit $S_{0}$ for each receiver above which we expect a detection rate of at least 50\,\%. For the 14\,mm receiver, we find $S_0=70.8^{+8.3}_{-6.3}$\,mJy, and for 7\,mm $S_{0}=337^{+36}_{-50}$\,mJy. For the 20\,mm receiver, we do not have enough data on the faintest sources to determine a flux density threshold for a detection rate of $50$\,\%. All sources with a 20\,mm flux density $\gtrsim30$\,mJy have a detection chance $>50$\,\%.

\begin{figure*}
    \centering
    \begin{subfigure}[b]{0.49\textwidth}
        \centering
        \includegraphics[width=\textwidth]{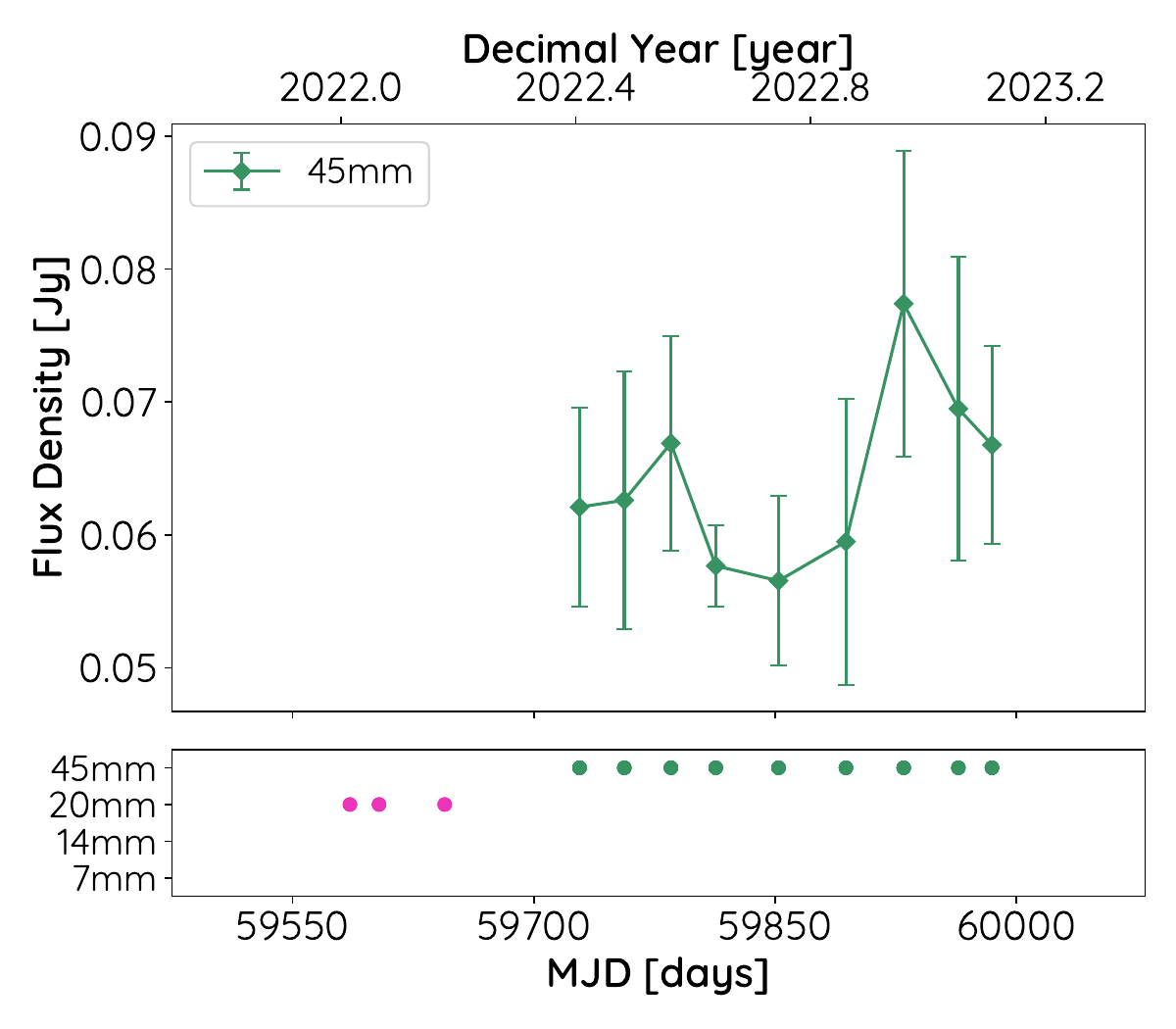}
        \caption[]%
        {{\small J0152+0146 (sub-sample I)}}    
        \label{fig:mean and std of net14}
    \end{subfigure}
    \hfill
    \begin{subfigure}[b]{0.49\textwidth}  
        \centering 
        \includegraphics[width=\textwidth]{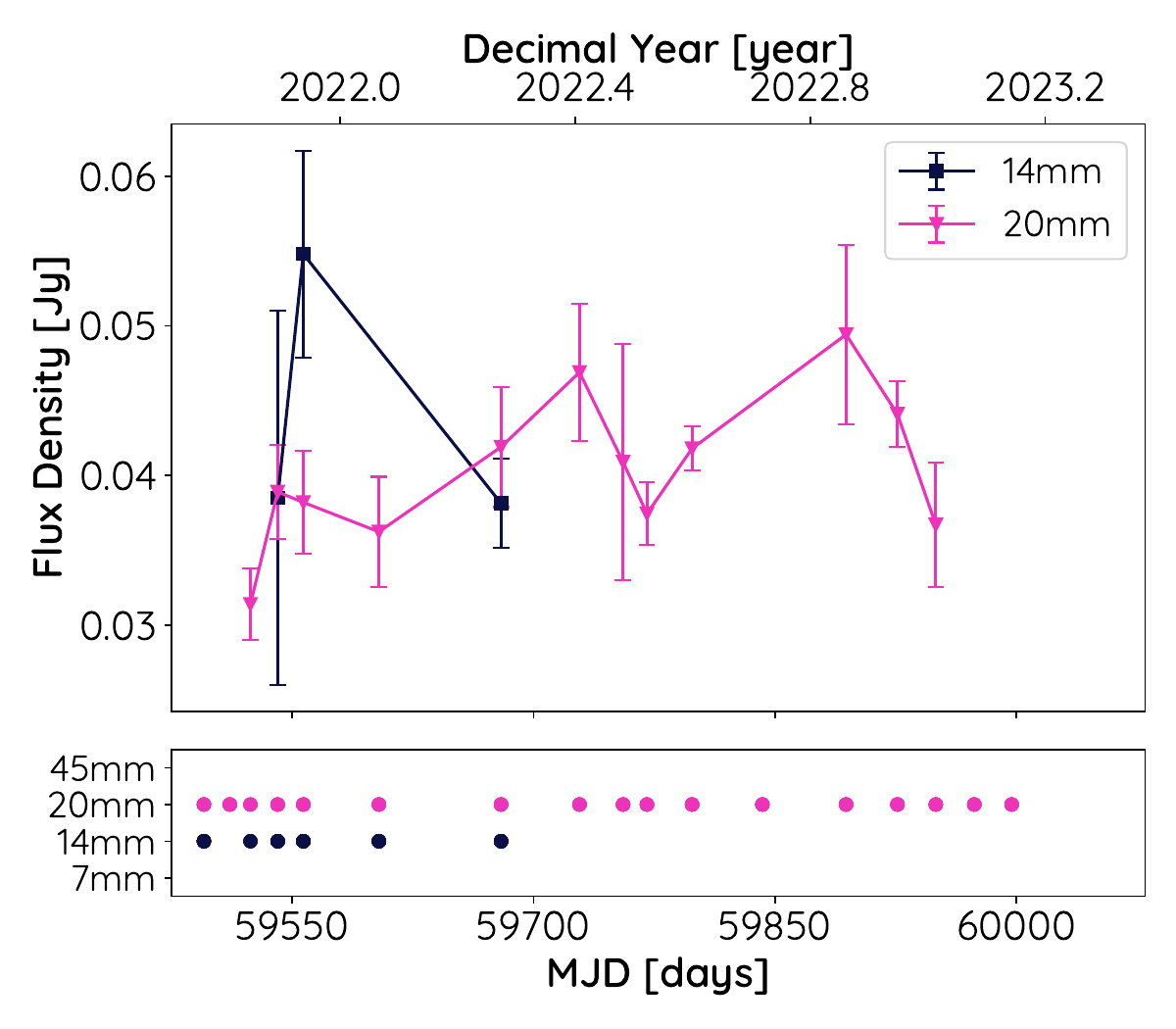}
        \caption[]%
        {{\small J0812+0237 (sub-sample II)}}    
        \label{fig:mean and std of net24}
    \end{subfigure}
    \vskip\baselineskip
    \begin{subfigure}[b]{0.49\textwidth}   
        \centering 
        \includegraphics[width=\textwidth]{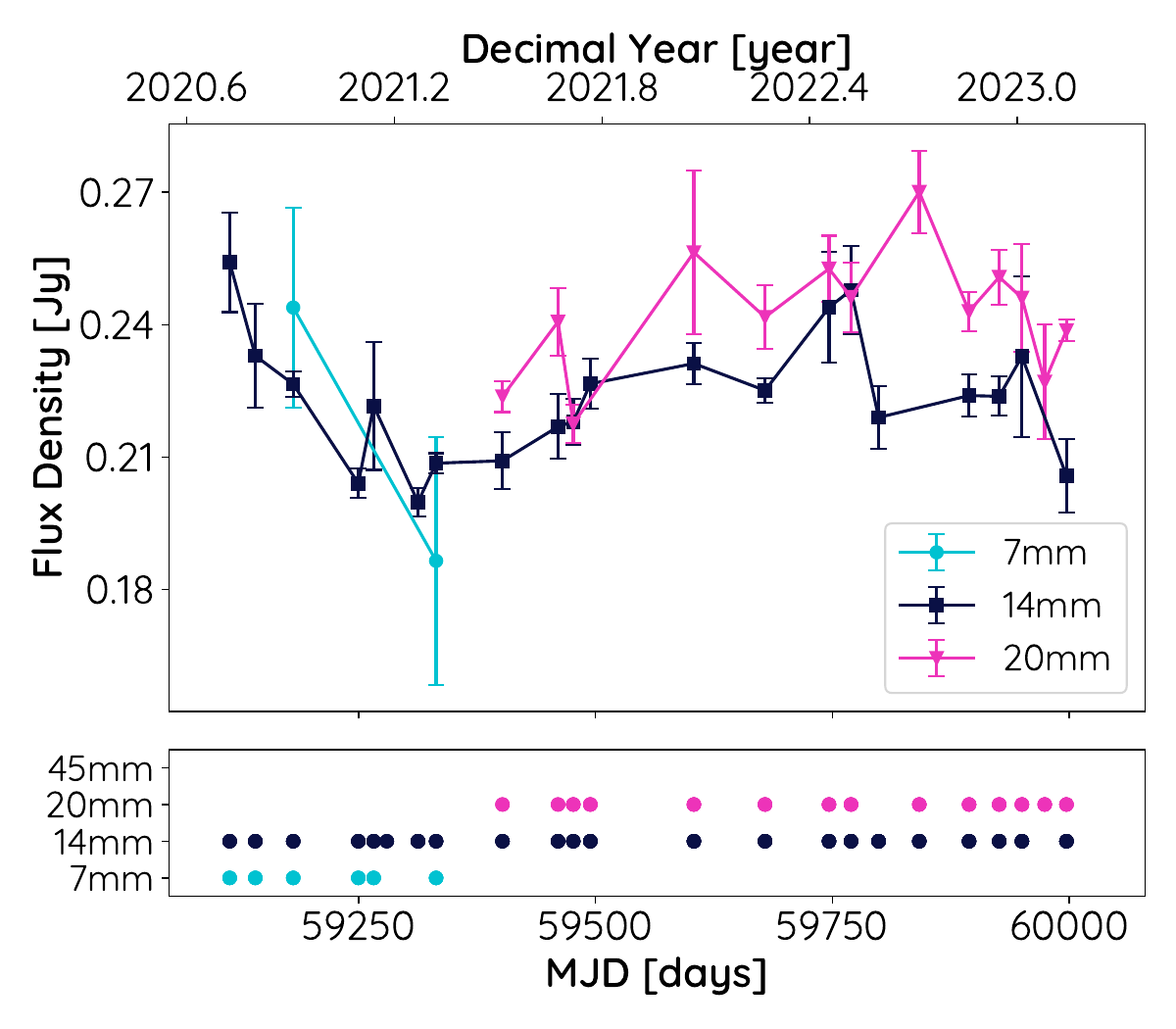}
        \caption[]%
        {{\small J1015+4926 (sub-sample III)}}    
        \label{fig:mean and std of net34}
    \end{subfigure}
    \hfill
    \begin{subfigure}[b]{0.49\textwidth}   
        \centering 
        \includegraphics[width=\textwidth]{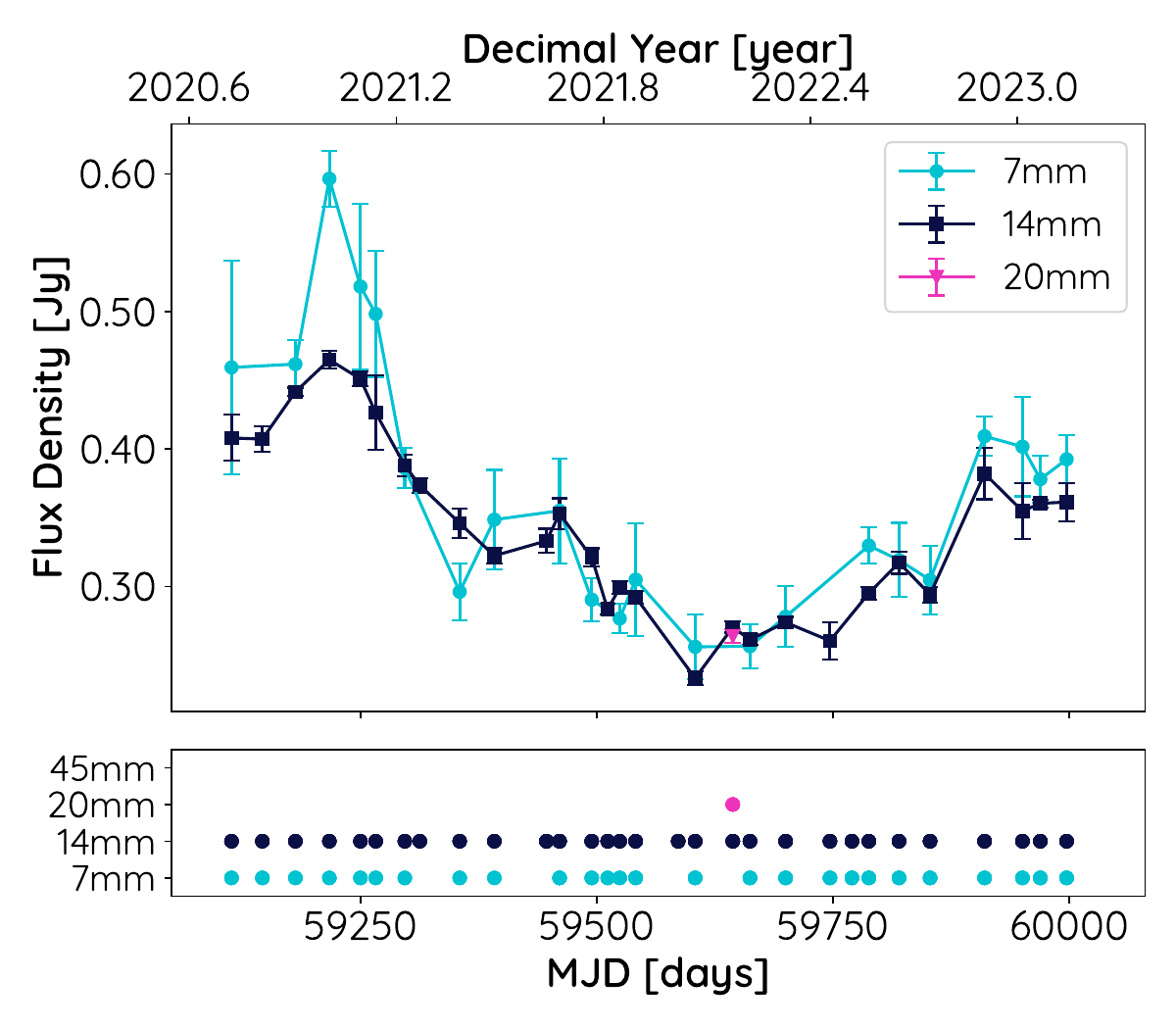}
        \caption[]%
        {{\small J1230+2518 (sub-sample IV)}}    
        \label{fig:mean and std of net44}
    \end{subfigure}
    \caption[]
    {\small Illustrative light curves (upper panels) of four selected sources from the TELAMON TeV-sample. The flux densities are averaged over the individual receiver sub-bands as described in Sect.\,\ref{sec:subbandaveraging}. The lower panels indicate the times of observation with the given receiver. If there is no matching flux density at the same time in the light curve, the source was not detected.} 
    \label{fig:example_lcs}
\end{figure*}

Sources below a 20\,mm flux density of $\sim$30\,mJy have a limited detection chance with either of the three introduced receivers. We therefore started observing these sources at even longer wavelengths (i.e., 45\,mm, sub-sample I) to have the best chances to detect them in our long-term monitoring program. The second faintest sources, exhibiting a 20\,mm flux density $\gtrsim30$\,mJy but a 14\,mm flux density $\lesssim70$\,mJy, are monitored at 20\,mm (sub-sample II) in the current observing setup. Brighter sources ($S_\mathrm{14mm}\gtrsim70\,\mathrm{mJy}$ and $S_\mathrm{7mm}\lesssim350$\,mJy) can be monitored at 20\,mm and 14\,mm (sub-sample III) in a sensible way, and for the brightest sources ($S_\mathrm{7mm}\gtrsim350$\,mJy) sensible monitoring is possible at both 14\,mm and 7\,mm (sub-sample IV). Note that in principle these thresholds can be improved by spending more time on each source, i.e., performing more sub-scans. However, this tactic is not favorable for our program, since we need to establish an ideal trade-off between observing time spent per source and the total number of sources that can be monitored. In Fig.\,\ref{fig:example_lcs}, we show example light curves of one representative source per sub-sample. One can see in light curves (a), (b), and (c), that we tried monitoring the sources with higher frequency receivers than suggested by the sub-sample, which resulted in fairly low detection rates. Since the adaption of the sub-samples, the detection rates have significantly improved. Ongoing TELAMON observation are conducted according to this strategy as it is the best possible setup for our purposes.

\section{Conclusions \& outlook} 
\label{sec:conclusions}
We have presented the first results of the pilot-phase (August 2020 - February 2023) of the TELAMON AGN monitoring program for a complete sample of all TeV-emitting blazars in the Northern Hemisphere. We have used the Effelsberg 100-m telescope to monitor these sources at high radio frequencies from 14\,GHz$-$44\,GHz every two to four weeks. We have developed a semi-automated data reduction pipeline which allows for immediate reduction of the data. From the observations, we derived flux densities and spectral indices from the sources in our sample. Plots of the latest available light curves and source spectra are publicly available on the dedicated TELAMON website\footnote{\url{https://telamon.astro.uni-wuerzburg.de/}}.

Our results are consistent with the findings of a prior study of the same object class by \cite{Lindfors}. In comparison to the GeV-selected F-GAMMA sample \citep{Fuhrmann,Angelakis2019}, we find that the spectral indices of both GeV- and TeV-selected sources are consistent with a flat spectrum. The average flux density at 14\,mm is significantly lower in the TeV-selected sample than what \cite{Angelakis2019} have found in their study.
The comparison between GeV- and TeV-emitting blazars is still limited by the small number of TeV-emitting blazars, which is expected to increase within the next years with upcoming observations from the Cherenkov Telescope Array \citep{CTA}.

In future studies, we will analyze the variability of TeV-blazars and perform correlation studies with multi-wavelength light curves, similar to previous studies of GeV-emitting AGN \citep[e.g.,][]{Paraschos,Roesch,Fuhrmann2014}. For several individual sources we have already demonstrated the multi-wavelength capabilities of our program \citep[e.g.,][]{Gokus2022, Raniere0658}, especially in combination with our complementary Southern Hemisphere monitoring program at the Australia Telescope Compact Array \citep[e.g.,][]{Eppel2023, Satalecka}. Moreover, the TELAMON program will continue its observations using an optimized observing setup including measurements at 45\,mm for very faint radio sources. In addition, we will provide results on follow-up observations of neutrino-candidate blazars, which have not been discussed in this paper. Another publication reporting on the polarization properties of our TeV-sample is currently in preparation \citep[cf. ][]{TELAMONpolarization}.

\begin{acknowledgements}
\review{We thank the anonymous referee for valuable comments that helped us improve the manuscript.}
This work is based on observations with the 100-m telescope of the MPIfR (Max-Planck-Institut für Radioastronomie) at Effelsberg. FE, SH, JH, MK and FR acknowledge support from the Deutsche Forschungsgemeinschaft (DFG, grants 447572188, 434448349, 465409577). We acknowledge the M2FINDERS project from the European Research Council (ERC) under the European Union's Horizon 2020 research and innovation programme (grant agreement No 101018682). \reviewtwo{PB is a member of the International Max Planck Research School for Astronomy and Astrophysics at the Universities of Bonn and Cologne.}
\end{acknowledgements}

%
%
%

\bibliographystyle{aa} 
\bibliography{bib} 

\begin{appendix}
\section{Flagging criteria for sub-scan fitting}
\label{app:flagging_criteria}

\begin{figure*}
    \centering
    \includegraphics[width=\textwidth]{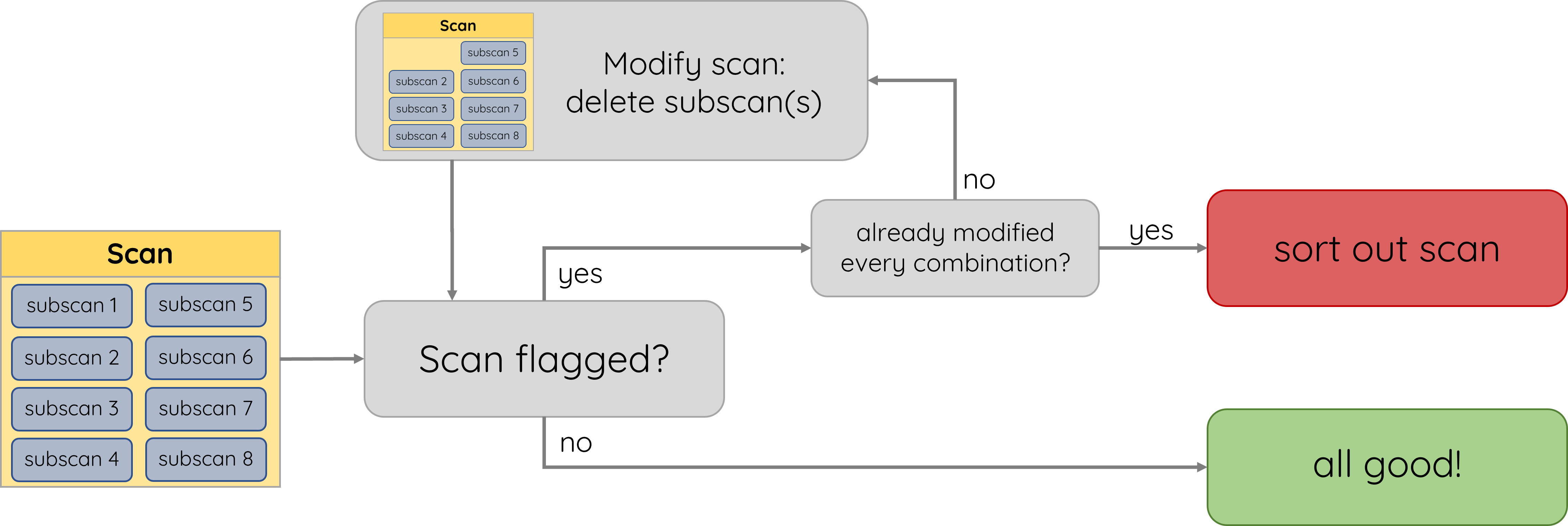}
    \caption{Flow diagram of the (semi-)automated flagging algorithm to sort out corrupted scans (cf. Sect.\ref{sec:subscan_fitting})}
    \label{fig:effpipe}
\end{figure*}

As mentioned in Sect.\,\ref{sec:subscan_fitting}, we apply different criteria to check for corrupted scans in our data analysis pipeline. The general data pipeline procedure is illustrated in Fig.\,\ref{fig:effpipe}. A scan is flagged if at least one of the following criteria applies:
\begin{itemize}
    \item The full-width at half maximum (FWHM) of the Gaussian fit of either the azimuth or elevation or both average scans deviates by more than $30\%$ from the frequency dependent half power beam width (HPBW). \textsl{Reason: The FWHM of a point-source scan is restricted by the angular resolution limit (beam width), depending on the observing frequency. Therefore, it should stay constant for each frequency and not vary significantly from source to source. There can be small positional telescope uncertainties throughout the sub-scans that lead to a slightly different FWHM of the averaged source scan, but these changes should not exceed $\sim30\%$. This flag type is typical for windy observing sessions, where the telescope's position changes slightly from sub-scan to sub-scan by wind gusts, or in cases where the source is below the detection limit and only background noise is fitted.} 
    \item The Gaussian fit of either azimuth or elevation or both average scans has negative amplitude. \textsl{Reason: The sources are expected to be point-like radio emitters, this means one expects to observe a Gaussian signal with a positive peak in total intensity. Since a negative amplitude indicates a dip in the Gaussian, there is something wrong with one or more sub-scans. It can also indicate that the source is below the detection limit and not visible at all, i.e., only background noise is fitted.} 
    \item The maximum of the Gaussian fit of either azimuth or elevation or both average scans has an offset from the scan center greater than $20\%$ of the HPBW. \textsl{Reason: The Gaussian fit is expected to be centered in azimuth and elevation scans. If this is not the case, the cross-scan has missed the source position by several arcseconds. In principle, these cases are corrected in the analysis (see Sect.~\ref{sec:pointcorr}), but at a level of more than $20\%$ of the beam width this offset correction is not accurate anymore.} 
    \item The amplitudes of the Gaussian fits of azimuth and elevation average scans differ by more than $15\%$. \textsl{Reason: The source is expected to be point-like and centered in azimuth and elevation scans. This means the average scans in azimuth and elevation should have similar amplitudes. If amplitudes differ by more than $15\%$ this is an indication of corrupted sub-scans included in the average scan or of the cross-scan not being centered at the source position.} 
\end{itemize}

Note that the exact limits that indicate when a scan is flagged rely on experience. The chosen values have proven to be well suited for the analysis in the used frequency bands. For every scan (usually consisting of 8 or 16 sub-scans), it is first checked if the averaged scan is flagged. If the scan is not flagged, it will be used as-is for the next analysis steps without performing any changes. If the scan is flagged, the algorithm tries to remove one sub-scan and checks if the average scan (excluding the removed sub-scan) is still flagged. If this new average scan is not flagged, the average scan, excluding the deleted sub-scan, will be used for further analysis steps. If the new average scan is still flagged this procedure goes on until the algorithm has found a combination of removed sub-scans that leads to a non-flagged average scan. The algorithm allows for up to two sub-scans in total to be removed from the initial scan. If the average scan is still flagged after trying to remove all sub-scan combinations, the scan is sorted out as a corrupted scan and it is not used for further analysis. 

\section{Results on additional TeV-sources}

    \begin{table*}[t]
    \caption{\small Average flux densities for all five Southern sources from the TELAMON sample and the TeV-detected radio galaxy J1145+1936 (3C\,264) as observed during the first 2.5 years. The median spectral index has been calculated across all receiver bands and epochs. Detection rates and the number of epochs each source was observed are shown.}
    \label{tab:average_table_south}
    \centering
    \footnotesize
    \begin{threeparttable}
    \begin{tabular}{@{}cc@{\,}ccccccccc@{}}
    \hline\hline
    ID  & $S_\textrm{20\,mm}$\tnote{a} & D-Rate\tnote{b} & N$_\mathrm{obs}$\tnote{c} & $S_\textrm{14\,mm}$\tnote{a} & D-Rate\tnote{b} & N$_\mathrm{obs}$\tnote{c} & $S_\textrm{7\,mm}$\tnote{a} & D-Rate\tnote{b} & N$_\mathrm{obs}$\tnote{c} & $\alpha$\tnote{d}\\
    (J2000) & [Jy] & [\%] & & [Jy] & [\%] & & [Jy] & [\%] & &\\
    \hline
    J0303$-$2407 & 0.201\,$\pm$\,0.015 & 90 & 10 & 0.185\,$\pm$\,0.023 & 88 & 18 & - & 0 & 7 & $-$0.38 \\
    J0913$-$2103 & 0.1459\,$\pm$\,0.0049 & 71 & 7 & 0.1201\,$\pm$\,0.0038 & 83 & 6 & - & - & 0 & $-$0.42 \\
    J1518$-$2731 & 0.220\,$\pm$\,0.032 & 80 & 5 & 0.207\,$\pm$\,0.020 & 77 & 9 & - & 0 & 3 & $-$0.30 \\
    J1958$-$3011 & 0.07635\,$\pm$\,0.00049 & 66 & 3 & 0.100\,$\pm$\,0.014 & 33 & 3 & - & - & 0 & $-$ \\
    J2158$-$3013 & 0.369\,$\pm$\,0.050 & 71 & 7 & 0.352\,$\pm$\,0.063 & 69 & 13 & - & 0 & 6 & $-$0.19 \\
    \hline
    J1145+1936 & 0.484\,$\pm$\,0.033 & 94 & 18 & 0.341\,$\pm$\,0.027 & 96 & 26 & 0.224\,$\pm$\,0.012 & 28 & 7 & $-$0.91 \\
    \hline
    \end{tabular}
    \begin{tablenotes}\footnotesize
    \tnote{a}Average flux density and standard deviation at the given wavelength taken over all observed epochs.
    \tnote{b}Detection rate (i.e., number of epochs with detection divided by the number of epochs where the source was observed) at the given wavelength.
    \tnote{c}Number of epochs where the source was observed at the given wavelength.
    \tnote{d}Median spectral index across all observed epochs with at least three sub-band detections.
    \end{tablenotes}
    \end{threeparttable}
    \end{table*}
    
\label{app:extra_sources}
For the statistical analysis in sections \ref{sec:results} and \ref{sec:discussion} we have excluded the five Southern TeV-blazars, which were monitored in the beginning of the program, and the radio galaxy J1145+1936 (3C\,264). Their average flux densities and 14\,mm average spectral indices are presented in Table\,\ref{tab:average_table_south}. For consistency, we have re-run all statistical tests that we performed in sections \ref{sec:results} and \ref{sec:discussion} including the five Southern TeV-blazars. We find that our results and conclusions are not significantly impacted. The similarity of F-GAMMA and TELAMON spectral indices becomes more significant ($p=0.19$), and the difference of HBLs and EHBLs within the TELAMON sample becomes slightly more significant at 20\,mm ($p=0.07$) and 14\,mm ($p=0.01$). We chose not to include these five sources in the main analysis since they are not chosen randomly and would introduce a statistical bias to the otherwise complete sample of all TeV-detected blazars in the Northern Hemisphere.

\end{appendix}

\end{document}